\documentclass[twocolumn,showpacs,amsfonts,aps,prc,floatfix,%
superscriptaddress]{revtex4} 

\usepackage{amsmath}
\usepackage{bm}
\usepackage{graphicx}

\newcommand{\Tdec}{T_\mathrm{dec}}
\newcommand{\edec}{e_\mathrm{dec}}

\newcommand{\be}[1]{\begin{equation}\label{#1}}
\newcommand{\ee}{\end{equation}}

\newcommand{\vtwopt}{$v_2(p_T)$}
\newcommand{\eq}{{\,=\,}}

\usepackage{color}

\begin{document}


\title{Mass ordering of differential elliptic flow and its violation
for $\phi$ mesons}
\date{\today}

\author{Tetsufumi Hirano}
\email[Correspond to\ ]{hirano@phys.s.u-tokyo.ac.jp}
\affiliation{Department of Physics, The University of Tokyo,
Tokyo 113-0033, Japan}

\author{Ulrich Heinz}
\affiliation{Department of Physics, The Ohio State University, 
Columbus, OH 43210, USA}
\affiliation{CERN, Physics Department, Theory Division, CH-1211 Geneva 23,
Switzerland}

\author{Dmitri Kharzeev}
\affiliation{Nuclear Theory Group, Physics Department, Brookhaven 
            National Laboratory, Upton, NY 11973-5000, USA}

\author{Roy Lacey}
\affiliation{Department of Chemistry, SUNY Stony Brook, Stony Brook, 
NY 11794-3400, USA}

\author{Yasushi Nara}
\affiliation{
Akita International University, 193-2 Okutsubakidai, Yuwa-Tsubakigawa, 
Akita 010-1211, Japan
}

\begin{abstract}
We simulate the dynamics of Au+Au collisions at the Relativistic Heavy Ion 
Collider (RHIC) with a hybrid model that treats the dense early quark-gluon 
plasma (QGP) stage macroscopically as an ideal fluid, but models the dilute 
late hadron resonance gas (HG) microscopically using a hadronic cascade. By
comparing with a pure hydrodynamic approach we identify effects of hadronic
viscosity on the transverse momentum spectra and differential elliptic flow
\vtwopt. We investigate the dynamical origins of the observed 
mass-ordering of \vtwopt\ for identified hadrons, focusing on 
dissipative effects during the late hadronic stage. We find that, at RHIC
energies, much of the finally observed mass-splitting is generated during 
the hadronic stage, due to build-up of additional radial flow. The $\phi$
meson, having a small interaction cross section, does not fully 
participate in this additional flow. As a result, it violates the 
mass-ordering pattern for $v_2(p_T)$ that is observed for other hadron 
species. We also show that the early decoupling of the $\phi$ meson
from the hadronic rescattering dynamics leads to interesting and 
unambiguous features in the $p_T$-dependence of the nuclear suppression
factor $R_{AA}$ and of the $\phi/p$ ratio.
\end{abstract}

\maketitle

\section{Introduction}
\label{sec1}
\vspace*{-2mm}

A presently hotly debated question is whether the quark-gluon plasma
(QGP) created in Au+Au collisions at the Relativistic Heavy Ion
Collider (RHIC) \cite{experiments} represents a ``perfect liquid''
\cite{reviews,HG05,Romatschke:2007mq,Song:2007fn}, i.e. a fluid
whose shear viscosity to entropy ratio $\eta/s$ is at or close to the 
conjectured \cite{Son_visc} lower bound  $\frac{\eta}{s}\eq\frac{1}{4\pi}$.
A key observable in this context is the elliptic flow $v_2$ of hadrons 
emitted anisotropically in non-central collisions \cite{Ollitrault}. At 
the highest RHIC energy of $\sqrt{s}=200\,A$\,GeV, the observed $v_2$ 
values near midrapidity ($|\eta|{\,\alt\,}1$), for not too large impact 
parameters ($b{\,\alt\,}7$\,fm) and transverse momenta 
($p_T{\,\alt\,}1.5$\,GeV/$c$), agree well with predictions from ideal 
fluid dynamics \cite{reviews} (i.e. assuming zero viscosity), including 
\cite{STARv2,PHENIXv2} the predicted dependence of $v_2$ on the 
transverse momentum $p_T$ and hadron masses \cite{HKHRV01}. 
Evidence for non-zero shear viscosity in the collision fireball is 
obtained from deviations from ideal fluid dynamical behavior. This is
manifested in the experimental data via a gradual break-down of the ideal 
fluid description when collisions are studied at larger impact parameters 
and at lower energies \cite{NA49v2sys} or when measurements are made away 
from midrapidity \cite{PHOBOSv2eta,Hiranov2eta,HT02}. In previous work 
\cite{HHKLN} we have shown that a large (and possibly the dominant) 
fraction of these deviations from ideal hydrodynamics is due to 
``late viscosity'', caused by dissipative effects in the dilute 
hadronic rescattering stage that stretches between hadronization of the 
QGP and final kinetic freeze-out. The question whether there is also 
non-negligible ``early viscosity'' in the dense QGP phase remains 
open. An answer to this question requires a proper viscous hydrodynamic 
treatment of the QGP fluid which is presently being pursued vigorously 
\cite{Romatschke:2007mq,Song:2007fn,visc_hydro}. It also depends on still 
unknown details of the initial conditions in heavy-ion collisions, in 
particular the initial spatial eccentricity of the fireball 
\cite{HHKLN,Drescher:2006pi,Lappi:2006xc,Drescher:2006ca}. 

In this paper we report additional results from the hybrid model study
presented in Ref.~\cite{HHKLN}, focusing our attention on a detailed 
investigation of dissipative effects during the late hadronic rescattering 
stage. The early QGP stage, including its hadronization, is described 
by ideal fluid dynamics. Specifically, we address the questions of (i)
how radial and elliptic flow evolve during the hadronic stage when 
it is described by a realistic hadronic rescattering cascade, rather 
than by an ideal fluid; (ii) how these differences affect the shapes 
of the finally observed hadronic $p_T$-spectra and their differential 
elliptic flow \vtwopt; and (iii) whether the differences between ideal 
fluid and realistic kinetic behavior during the late hadronic stage are 
similar for all hadronic species, or whether different magnitudes of 
their scattering cross sections translate into measurably different 
characteristics of their observed spectra and elliptic flow.

The paper is organized as follows: For completeness, we begin in 
Section~\ref{sec2} with a short review of the hybrid model \cite{HHKLN} 
employed in this study. Our results are presented in Section~\ref{sec3}, 
in three subsections organized along the questions raised in the preceding
paragraph. We close the paper by presenting our conclusions and some 
perspectives in Section~\ref{sec4}.  
  
\section{The model}
\label{sec2}

Our study is based on a hybrid model which combines an ideal fluid 
dynamical description of the QGP stage with a realistic kinetic 
simulation of the hadronic stage \cite{HHKLN}. Relativistic hydrodynamics 
is the most relevant formalism to understand the bulk and transport 
properties of the QGP since it directly connects the collective flow 
pattern developed during the QGP stage with its equation of state (EOS).
It is based on the key assumption of local thermalization. Since this
assumption breaks down during both the very anisotropic initial matter 
formation stage and the dilute late hadronic rescattering stage, the
hydrodynamic model can be applied only during the intermediate period,
between initial thermalization after about 1\,fm/$c$ \cite{reviews}
and the completion of the quark-hadron transition which, in central 
Au+Au collisions at RHIC energies, happens after about 10\,fm/$c$.  

In absence of a non-equilibrium dynamical model for the early 
pre-equilibrium stage of the collision, its output is replaced by
a set of initial conditions for the hydrodynamic evolution which are
tuned to experimental measurements of the final state in central ($b\eq0$)
collisions \cite{reviews}. To describe the breakdown of the hydrodynamical
model during the late hadronic stage, due to expansion and dilution of the 
matter, one has two options: One can either impose a sudden transition
from thermalized matter to non-interacting, free-streaming hadrons through
the Cooper-Frye prescription \cite{CF}, imposed at a suitable value of
the {\em decoupling temperature} $\Tdec$ or {\em decoupling energy density} 
$\edec$ \cite{reviews}, or make a transition from the macroscopic 
hydrodynamic description to a microscopic kinetic description at a suitable 
value for the {\em switching temperature} $T_\mathrm{sw}$ where both 
descriptions are simultaneously valid 
\cite{BassDumitru,TLS01,NonakaBass,HHKLN,fn1}, letting the subsequent
kinetic decoupling play itself out automatically by following the 
microscopic evolution until all interactions have ceased. We here use
both approaches alternatively, in order to isolate effects that are
specifically caused by dissipative effects in the hadron rescattering 
cascade. 

\subsection{Ideal Hydrodynamics}
\label{sec2a}

For the space-time evolution of the perfect QGP fluid we solve numerically
the equations of motion of ideal fluid dynamics, for a given initial state,
in three spatial dimensions and in time ((3+1)-d ideal hydrodynamics) 
\cite{Hiranov2eta}:
\begin{eqnarray}
\label{eq:hydro}
\partial_\mu T^{\mu \nu} & = & 0, \\
\label{eq:tmunu}
T^{\mu \nu} &  =  &(e+p) u^\mu u^\nu -p g^{\mu \nu}.
\end{eqnarray}
Here $e$, $p$, and $u^\mu$ are energy density, pressure, and four-velocity
of the fluid, respectively. Due to its smallness at collider energies, we 
neglect the net baryon density. As an algorithm to solve the above equations
we choose the Piecewise Parabolic Method (PPM) \cite{PPM}. It is known to be
a very robust scheme for solving non-relativistic gas dynamics including
shock wave formation and has been employed in many fields. We first applied 
it in \cite{Hirano:2000eu} to solve Eulerian hydrodynamics for relativistic 
heavy-ion collisions, Eqs.~(\ref{eq:hydro}) and (\ref{eq:tmunu}). Use of 
this algorithm enables us to describe the space-time evolution of 
relativistic fluids accurately even if the matter passes through a 
first-order phase transition. The PPM is a higher order extension
of the piecewise linear method employed, for example, in the rHLLE algorithm 
\cite{rhlle}. We solve Eq.~(\ref{eq:hydro}) in $(\tau,\,x,\,y,\,\eta_s)$
coordinates \cite{Hiranov2eta} where $\tau\eq\sqrt{t^2{-}z^2}$) and 
$\eta_s\eq\frac{1}{2}\ln[(t{+}z)/(t{-}z)]$ are longitudinal proper time 
and space-time rapidity, respectively, adequate for the description of 
collisions at ultra-relativistic energies. The grid sizes are 
$\Delta\tau\eq0.3$\,fm/$c$, $\Delta x\eq\Delta y\eq0.3$\,fm, and 
$\Delta\eta_s\eq0.3$. We have checked the grid size dependence of 
our final results and observed sufficient convergence with the given 
choice of grid parameters, as long as smooth initial conditions
such as those discussed below are used.

\subsection{Equation of State}
\label{sec2b}

For the high temperature ($T>T_c=170$ MeV) QGP phase we use the
EOS of massless non-interacting parton gas 
($u$, $d$, $s$ quarks and gluons) with a bag pressure $B$:
\begin{eqnarray}
  p = \frac{1}{3}(e-4B)
\end{eqnarray}
The bag constant is tuned to $B^{\frac{1}{4}} = 247.19$\,MeV to ensure
a first order phase transition to a hadron resonance gas at critical
temperature $T_c\eq170$\,MeV. The hadron resonance gas model at
$T<T_c$ includes all hadrons up to the mass of the $\Delta(1232)$
resonance. Systematic studies with various models of the EOS
including a more realistic cross-over one
will be discussed elsewhere.

For a meaningful discrimination between the ideal fluid and hadron 
cascade descriptions of the hadron phase, and a realistic direct 
comparison of hydrodynamic results with experimental data, our hadron 
resonance gas EOS implements chemical freeze-out at 
$T_\mathrm{chem}{\eq}T_c\eq170$\,MeV, as observed in RHIC collisions 
\cite{BMRS01}. This is achieved by introducing appropriate 
temperature-dependent chemical potentials $\mu_i(T)$ for all hadronic 
species $i$ in such a way that their numbers $\tilde N_i$ including 
all decay contributions from higher-lying resonances, $\tilde{N}_i 
= N_i + \sum_R b_{R\rightarrow i X}N_{R}$, are conserved during the 
evolution \cite{HT02,Bebie,spherio,Teaney:2002aj,Kolb:2002ve,Huovinen:2007xh}. 
[Here $N_i$ is the number of the $i$-th hadron, and $b_{R\rightarrow i X}$ 
is the effective branching ratio (a product of branching ratio and 
degeneracy) of a decay process $R\rightarrow i+X$.] In this ``PCE model'' 
\cite{HT02} only strongly interacting resonances with large 
decay widths (whose decays do not alter $\tilde N_i$) remain 
chemically equilibrated below the chemical freeze-out temperature.

The hadronic chemical composition described by hydrodynamics using the 
PCE model EOS is roughly consistent with that of the hadronic cascade 
models, as long as the latter are initialized at $T_{\mathrm{sw}}$ with 
thermal and 
chemical equilibrium distributions \cite{fn2}. This is crucial for a 
meaningful comparison between hydrodynamic and kinetic descriptions of 
hadronic matter since the chemical composition of the hadron resonance gas 
has a significant influence on the hydrodynamic evolution of the hadronic 
transverse momentum spectra \cite{HG05}: While the non-equilibrium hadronic
chemical potentials $\mu_i(T)$ do not affect the EOS $p(e)$ of the hadronic 
phase \cite{HT02}, and thus lead to almost identical evolution of
radial flow and total momentum anisotropy as for a chemically equilibrated 
hadron gas, they significantly alter the relationship between energy 
density and temperature, leading to cooler temperatures and hence to
steeper transverse momentum spectra at identical kinetic decoupling energy 
densities \cite{HT02}. This effect is seen most dramatically
in the time-dependence of the mean transverse momentum for pions 
\cite{HG05}: $\langle p_T\rangle_\pi$ \textit{decreases} with proper 
time after chemical freeze-out whereas with continued hadronic chemical 
equilibrium it would \textit{increase} with time. Clear conclusions about
hadronic dissipative effects on the shapes of the transverse momentum 
spectra can therefore only be drawn from a comparison with hydrodynamic
models that implement chemical and kinetic freeze-out separately.

\subsection{Initial Conditions}
\label{sec2c}

Contrary to Ref.~\cite{HHKLN} where we studied both Glauber model and 
Color Glass Condensate (CGC) type initial conditions, for the comparative 
study presented here we concentrate on the Glauber model, suitably generalized 
to account for the longitudinal structure of particle multiplicity 
\cite{HHKLN,AG05}. We assume an initial entropy distribution of massless 
partons according to
\begin{eqnarray}
\label{eq:dSdx}
 \frac{dS}{d\eta_s d^2x_\perp} &=& 
 {C\over 1+\alpha}\, \theta\bigl(Y_b{-}|\eta_s|\bigr)\, f^{pp}(\eta_s)
\nonumber\\
 &\times& \Biggl[\alpha 
  \left(\frac{Y_b{-}\eta_s}{Y_b}\,\frac{dN^A_{\rm part}}{d^2x_\perp} 
      + \frac{Y_b{+}\eta_s}{Y_b}\,\frac{dN^B_{\rm part}}{d^2x_\perp}\right)
\nonumber\\
 && \  + (1{-}\alpha)\, \frac{dN_{\rm coll}}{d^2x_\perp}\Biggr],
\end{eqnarray}
where $\bm{x}_\perp\eq(x,y)$ is the position perpendicular to the 
beam axis, $C$ is a normalization factor, the ``soft fraction'' $\alpha$ 
is explained below, the parameter $Y_b$ is the beam rapidity, and 
$f^{pp}$ is a suitable parametrization of the shape of rapidity 
distribution in $pp$ collisions:
\begin{equation}
\label{eq:pp}
  f^{pp}(\eta_s) = \exp\left[-\theta(|\eta_s|{-}\Delta\eta)\,
  \frac{(|\eta_s|{-}\Delta\eta)^2}{\sigma_\eta^2}\right].
\end{equation}
We study Au+Au collisions at $\sqrt{s}\eq200\,A$\,GeV and use $C\eq24$, 
$\Delta\eta\eq1.3$, and $\sigma_\eta\eq2.1$, so chosen as to reproduce 
the charged hadron pseudorapidity distributions measured in these collisions
\cite{PHOBOS_dNdeta}. $N^{A,B}_{\rm part}$ and $N_{\rm coll}$ are the 
number of wounded nucleons in each of the two nuclei and the number
of binary nucleon-nucleon collisions, respectively. These are calculated
from the Glauber model nuclear thickness function $T_{A,B}(\bm{x}_\perp)$
\cite{Kolb:2001qz}:
\begin{eqnarray}
\label{eq:3}
\frac{dN^A_{\rm part}}{d^2x_\perp} &=& 
T_A(r_+)\left[1 - 
\left(1-\frac{\sigma_{NN}^{\rm in}\,T_B(r_-)}{B}\right)^B
        \right],
\\  
\label{eq:4}
\frac{dN^B_{\rm part}}{d^2x_\perp} &=& 
T_B(r_-)\left[1 - \left(1-\frac{\sigma_{NN}^{\rm in}\,T_A(r_+)}{A}\right)^A
        \right],
\\  
\label{eq:5}
\frac{dN_{\rm coll}}{d^2x_\perp} &=& 
\sigma_{NN}^{\rm in}\,T_A(r_+)\,T_B(r_-).
\end{eqnarray}
Here $\sigma_{NN}^{\rm in} = 42$\,mb is the inelastic nucleon-nucleon 
cross section, and 
$r_\pm\eq\bigl[\left(x{\pm}\frac{1}{2}b\right)^2+y^2\bigr]^{1/2}$ 
(where $b$ is the impact parameter). 

The soft/hard fraction $\alpha\eq0.85$ was adjusted to reproduce the 
measured centrality dependence \cite{PHOBOS_Nch} of the charged hadron 
multiplicity at midrapidity. At $\eta_s\eq0$, Eq.~(\ref{eq:dSdx}) reduces 
to $\frac{dS}{d\eta_s d^2x_\perp} \propto
\frac{1}{1{+}\alpha} \bigl[\alpha \bigl(n^A_{\rm part}{+}n^B_{\rm part}\bigr) 
+ (1{-}\alpha)n_{\rm coll}\bigr]$ where $n{\,\equiv\,}\frac{dN}{d^2x_\perp}$
\cite{KH05}; this parameterization is equivalent to the one used in 
\cite{KLN01}, $\propto \bigl[\frac{1{-}x}{2} \bigl(n^A_{\rm part}{+}n^B_{\rm 
part}\bigr) + x n_{\rm coll}\bigr]$, with $x\eq\frac{1{-}\alpha}{1{+}\alpha}$. 
From Eq.~(\ref{eq:dSdx}), we can compute the entropy density at the initial 
time $\tau_0\eq0.6$\,fm/$c$ \cite{reviews} of the hydrodynamic evolution, 
$s(\tau_0,\bm{x}_\perp, \eta_s)\eq\frac{dS}{\tau_0 d\eta_s d^2x_\perp}$,
which provides the initial energy density and pressure distributions 
through the tabulated EOS described above.

Glauber model initial conditions have a long tradition for hydrodynamic 
simulations of heavy-ion collisions. In our previous study \cite{HHKLN}
we showed that with such initial conditions ``late viscosity'' effects
during the dilute hadronic rescattering stage are sufficient to explain
all observed deviations of elliptic flow measurements from ideal fluid
dynamical predictions. No significant additional viscous effects during
the early QGP stage were necessary. We also noted, however, that this
conclusion depends crucially on this particular choice of initial 
conditions, specifically the initial source eccentricity predicted by
the Glauber model. The good agreement between theory and experiment
disappears when one instead calculates the initial conditions from the
KLN model \cite{KLN01,Hirano:2004rs,Kuhlman:2006qp,Drescher:2006pi,%
Lappi:2006xc,Drescher:2006ca}, which is based on 
CGC ideas and, for the same impact parameter, produces almost 30\% 
larger source eccentricities. If Nature gives preference to such more
eccentric initial conditions, additional viscous effects and/or a 
softer EOS for the QGP stage may be needed to reproduce the experimental
data \cite{HHKLN,Drescher:2007cd}. Here, we will not pursue this line of 
thought any further, but focus on the case of Glauber model initial 
conditions and the specific modifications of hadron spectra and flow 
caused by ``late hadronic viscosity''. 

\subsection{Hadronic Cascade Model}
\label{sec2d}

In our hybrid model simulations we switch from ideal hydrodynamics to 
a hadronic cascade model at the switching temperature 
$T_{\mathrm{sw}}\eq169$\,MeV. The subsequent hadronic rescattering 
cascade is modeled by JAM \cite{jam}, initialized with hadrons 
distributed according to the hydrodynamic model output, calculated
with the Cooper-Frye formula \cite{CF} along the $T_{\mathrm{sw}}\eq169$\,MeV 
hypersurface rejecting inward-going particles. We have checked \cite{HHKLN}
that switching from an ideal hydrodynamic to a hybrid model description
does not entail a major readjustment of initial conditions: Keeping
the same initial conditions and hard/soft fraction $\alpha$ as previously 
determined within a purely hydrodynamic approach (see 
\cite{reviews,HT02} for a detailed discussion of that procedure) we 
find \cite{HHKLN} that the centrality dependence of $dN_{\mathrm{ch}}/d\eta$
at midrapidity remains consistent with the experimental data even if
we switch below $T_{\mathrm{sw}}$ to the hadronic cascade. Effects on 
the hadron spectra and elliptic flow are significant, however, and will
be discussed in the next section.

As customary in hadronic cascade models \cite{jam,RQMD,UrQMD}, JAM 
implements experimental hadronic scattering cross section data where 
available and uses the additive quark model where data do not exist,
assuming the following formula for the total cross section:
\begin{eqnarray}
\label{eq:aqm}
  \sigma_{\mathrm{tot}} & = &
  \sigma_{NN}^{\mathrm{tot}}\frac{n_1}{3}\frac{n_2}{3} 
  \left(1-0.4\frac{n_{s1}}{n_1}\right)\left(1-0.4\frac{n_{s2}}{n_2}\right).
\end{eqnarray}
Here $\sigma_{NN}^{\mathrm{tot}}$ is the total nucleon-nucleon cross 
section, $n_{i}$ is the number of constituent quarks in a hadron, and
$n_{si}$ is the number of strange quarks in a hadron. For hadrons 
composed entirely of strange quarks, such as $\phi\eq(s\bar{s})$ and 
$\Omega\eq(sss)$, the cross sections become very small, due to the 
suppression factors in brackets in Eq.~(\ref{eq:aqm}). Only when we 
calculate spectra for $\phi$ mesons in Sec.~\ref{sec3c}, the decay 
channels for $\phi$ mesons are switched off in the hadronic cascade 
calculations. Since the life time of $\phi$ mesons ($\approx 46$ fm/$c$) 
is longer than the typical life time of the system ($\sim 10$-20 fm/$c$),
and the number of $\phi$ mesons is small compared to pions, kaons, and
nucleons, this prescription is not expected to affect the bulk 
space-time evolution during the hadronic stage.

\section{Results}
\label{sec3}

In Ref.~\cite{HHKLN} we investigated the effect of hadronic dissipation 
on elliptic flow and found that it significantly suppresses the 
$p_{T}$-integrated $v_2$ at forward and backward rapidity and in
peripheral collisions. In the following we explore the origins of
this finding in more detail, by investigating hadronic dissipative
effects on hadron spectra and differential elliptic flow \vtwopt.
We finally explore specifically the spectra and elliptic flow of 
$\phi$ mesons as an example of a hadron that is only weakly coupled
to the rest of the expanding hadronic fireball.

\subsection{Hadronic dissipative effects on spectra and elliptic flow}
\label{sec3a}

In this subsection, we compare results from the hybrid model with the 
ones from ideal hydrodynamics. In ideal hydrodynamic calculations it is
assumed that even the late hadron resonance gas phase is characterized
by essentially vanishing mean free paths and thus behaves as a perfect 
fluid, all the way down to kinetic decoupling of the hadron momenta
at $T_{\mathrm{th}}\eq100$\,MeV. (This value is obtained by a simultaneous
fit of the pion and proton spectra in central collisions which allows to 
separate the effects of radial flow and thermal motion at kinetic
freeze-out \cite{reviews}.) As discussed, chemical freeze-out is 
implemented at $T_\mathrm{chem}\eq170$\,MeV by using an EOS with 
non-equilibrium chemical potentials which hold the stable particle 
yields constant (and close to the ones in the cascade model approach) 
during the hydrodynamic evolution of the hadronic phase. The key 
difference between the hydrodynamic and hybrid model approaches is,
thus, the finite mean free path for momentum-changing collisions in 
the hadronic cascade. 
%
 \begin{figure}[htb]
\includegraphics[width=3.4in]{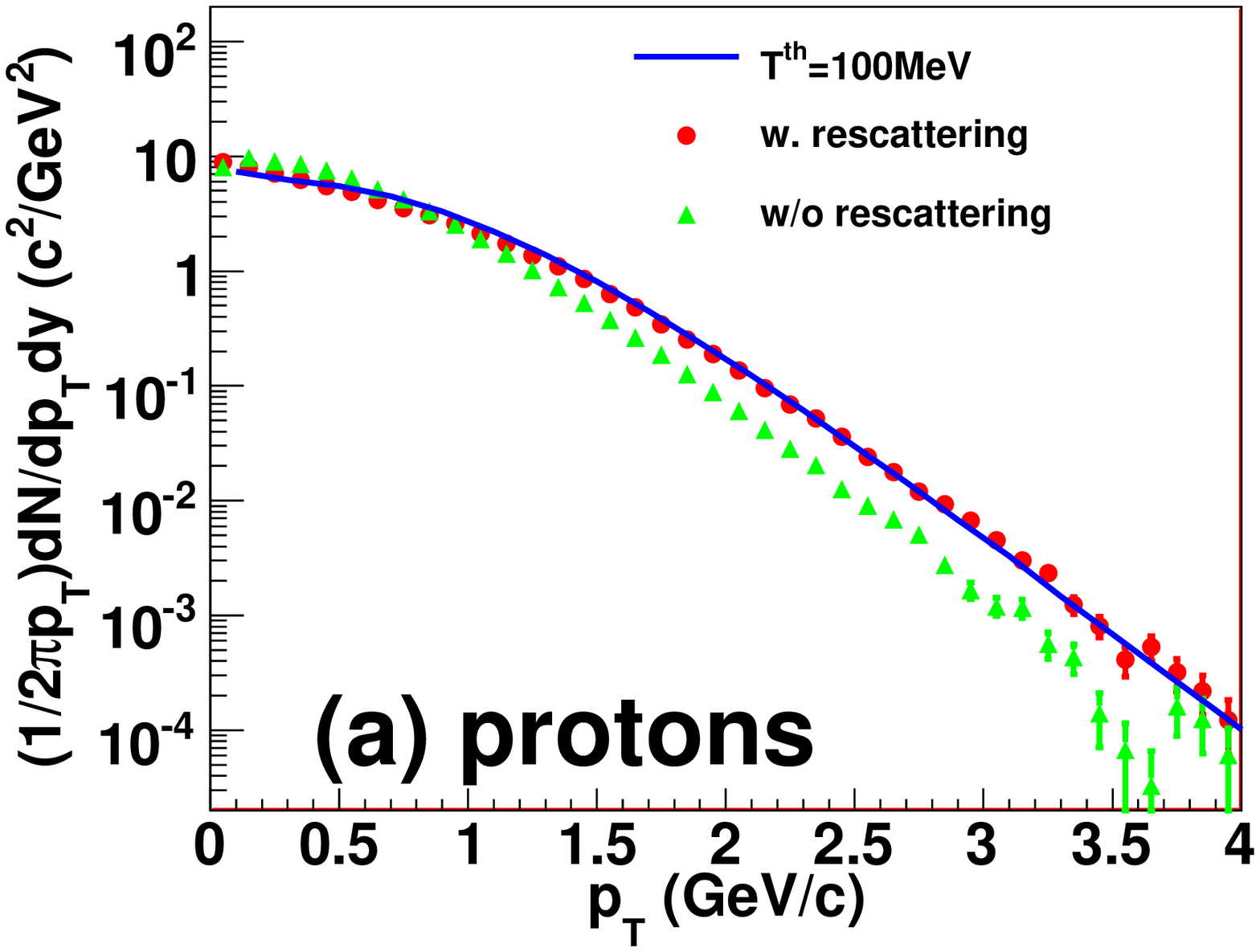}
\includegraphics[width=3.4in]{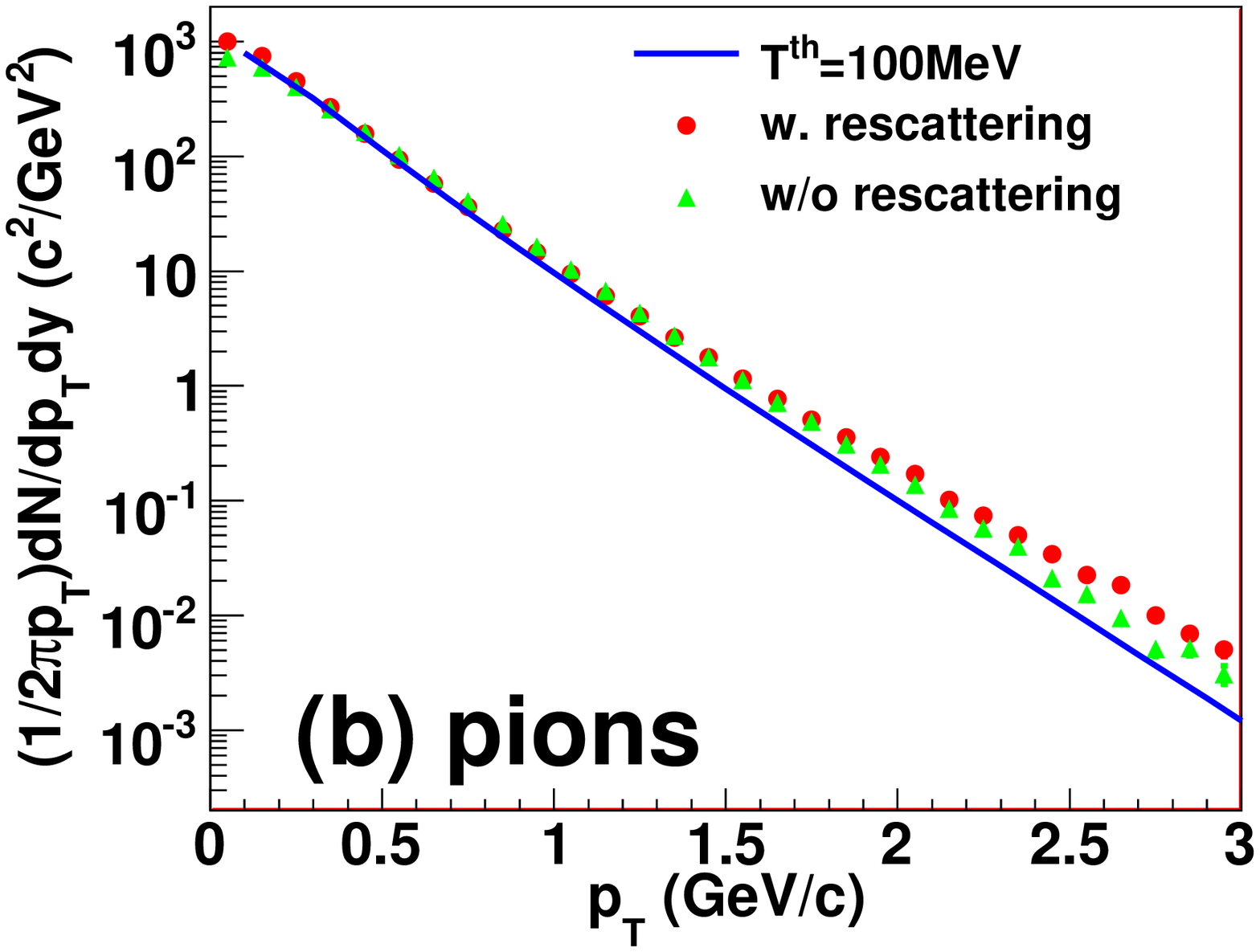}
\caption{(Color online)
  $p_{T}$ spectra with and without hadronic rescattering for (a) protons 
  and (b) pions at midrapidity for Au+Au collisions at $b\eq2$\,fm,
  compared with results from ideal hydrodynamics decoupling  at
  $T_{\mathrm{th}}\eq100$\,MeV.
}
 \label{fig:dndpt_hydro}
 \end{figure}
%

In Fig.~\ref{fig:dndpt_hydro}, $p_{T}$-spectra for protons and pions are 
shown for both, the hybrid model and the ideal hydrodynamic approach.
For comparison, we also plot the $p_{T}$-spectra without hadronic 
rescattering, obtained by setting all cross sections to zero in the
hadron cascade or by setting $T_\mathrm{th}\eq{T}_\mathrm{sw}\eq169$\,MeV
in the hydrodynamic approach (both procedures give the same spectra,
by construction). Note here that we include contributions
from all resonances (except for weak decays unless explicitly noted
otherwise) in
ideal hydrodynamic and hybrid-model results. 
One sees that hadronic rescattering in the JAM cascade
pushes the protons to higher $p_{T}$ in exactly the same way as the 
growing radial flow does in the hydrodynamic approach, if one chooses
for the latter a kinetic decoupling temperature of 
$T_{\mathrm{th}}\eq100$\,MeV. The reasonable fit of the measured
proton $p_T$-spectra \cite{phenix:pid200} up to $p_T{\,\sim\,}1.5$\,GeV/$c$
by the hydrodynamic model \cite{reviews,HT02,Kolb:2002ve} thus
persists in the hybrid model approach (see Fig.~\ref{fig:dndpt} in the 
following subsection).

The lack of visible dissipative effects on the proton spectra is
probably an artifact caused by a judicial choice of the kinetic freeze-out
temperature $T_{\mathrm{th}}\eq100$\,MeV in the hydrodynamic approach,
which was driven by the wish to reproduce the measured proton spectra
with this model. This accident does not repeat itself for the pions, shown 
in Fig.~\ref{fig:dndpt_hydro}(b). For pions, the $p_T$-spectrum becomes 
slightly steeper when evolved hydrodynamically (the steepening effects
due to cooling are not quite compensated by the increasing radial flow)
whereas it gets hardened by hadronic rescattering effects in the hybrid 
approach. 

This pattern is consistent with theoretical expectations: In the ideal 
fluid approach, $pdV$ work in the longitudinal direction reduces the 
transverse energy per unit rapidity 
\cite{Gyulassy:1983ub,Ruuskanen:1984wv}. Since pions dominate
the medium but their number is fixed after chemical freeze-out, this 
leads to a decrease of the average $p_T$ per pion \cite{HG05}, explaining 
the steeper pion spectrum from ideal hydrodynamics. (This argument is not 
quantitative since it neglects the shifting balance of transverse energy
carried by pions and heavier particles such as protons which are more
strongly affected by the developing radial flow \cite{HG05}. Also note 
that it does not remain true if a chemical equilibrium EOS is used in 
the hadronic phase where the pion number decreases with temperature and 
the average transverse energy per pion thus increases \cite{HG05}.)
In contrast to the ideal fluid, the hadron gas in the JAM cascade is
highly viscous. Shear viscosity is known to reduce the longitudinal 
and increase the transverse pressure \cite{visc_hydro}, reducing the
loss of transverse energy due to longitudinal $pdV$ work and increasing
the transverse flow due to larger transverse pressure gradients
\cite{visc_hydro}. In addition, there are viscous corrections to the
(flow-boosted) thermal equilibrium form of the distribution function
at kinetic freeze-out which lead to an additional viscous distortion
of the $p_T$-spectrum which actually increases with $p_T^2$ 
\cite{Teaney_visc}. For Bjorken expansion of a homogeneous cylinder
this distortion can be written analytically as \cite{Teaney_visc}
\begin{eqnarray}
  \frac{dN}{p_{T}dp_{T}} 
  \approx \left(1+\frac{\Gamma_s}{4 \tau_f T^2} p_{T}^2 \right)
  \frac{dN_{0}}{p_{T}dp_{T}}
\label{eq:visccorr}
\end{eqnarray}
where $\frac{dN_0}{p_Tdp_T}$ is the spectrum calculated from a 
boosted thermal equilibrium distribution along the decoupling surface 
at freeze-out time $\tau_f$ and temperature $T$, and the expression in 
brackets preceding it is the $p_T^2$-dependent viscous correction, 
parametrized by the sound attenuation length 
$\Gamma_s\eq\frac{4}{3}\frac{\eta}{sT}$ (where $\eta$ is the shear viscosity).

The viscous flattening of the pion spectrum relative to the pure 
hydrodynamic approach seen in Fig.~\ref{fig:dndpt_hydro}(b) receives 
contributions from both factors in Eq.~(\ref{eq:visccorr}): 
$\frac{dN_0}{p_Tdp_T}$ is flattened by the larger transverse flow 
generated by the viscously increased transverse pressure, and 
additional flattening comes from the factor in brackets, due to a 
non-zero value for $\Gamma_s$ in a viscous fluid. We don't know which 
of the two effects is larger;
we only note that the pion spectrum from the hybrid model can be fitted 
very well by simply multiplying the hydrodynamic model spectrum 
with the factor in brackets in Eq.~(\ref{eq:visccorr}), taking 
$T\eq{T}_\mathrm{th}\eq100$\,MeV and adjusting $\Gamma_s/\tau_f\eq0.01$. 
How meaningful such a fit is (given that the form (\ref{eq:visccorr})
makes unrealistic assumptions about the fireball expansion) remains to
be seen when realistic viscous hydrodynamic studies become available.

%
 \begin{figure}[htb]
 \includegraphics[width = \linewidth,clip]{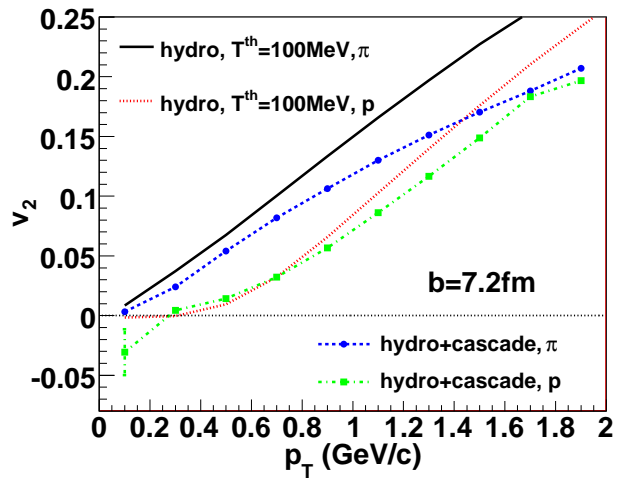}
 \caption{(Color online)
   $v_2(p_{T})$ for pions and protons in $|\eta|{\,<\,}1.3$ at $b=7.2$ fm.
   Results for pions (solid) and protons (dotted) from ideal hydrodynamics
   with $T_{\mathrm{th}}\eq100$\,MeV are compared with the ones for pions 
   (dashed) and protons (dash-dotted) from the hybrid model.
 }
 \label{fig:v2pt_hydro}
 \end{figure}
%

%
\begin{figure*}[bht]
\includegraphics[width=2.3in]{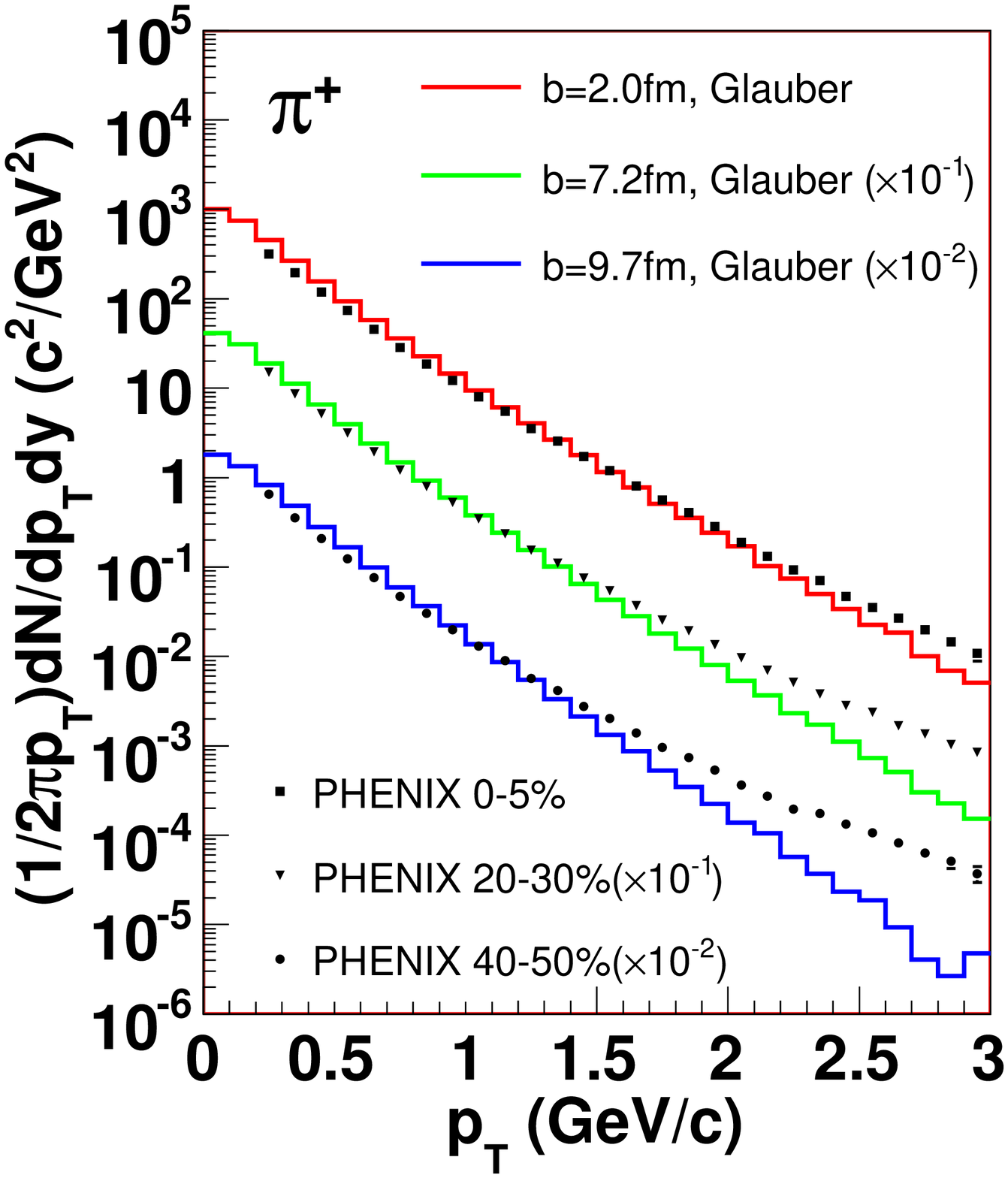}
\includegraphics[width=2.3in]{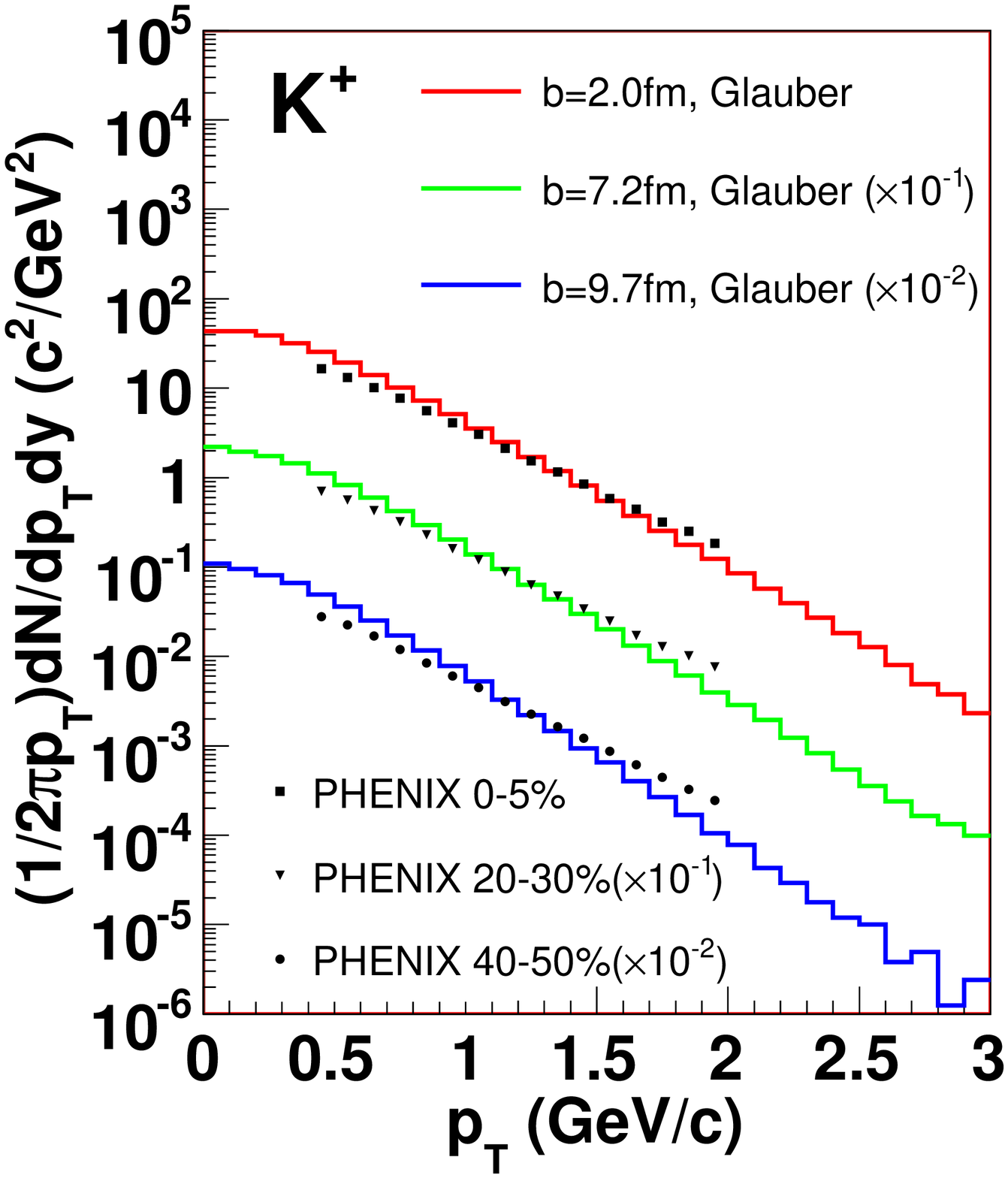}
\includegraphics[width=2.3in]{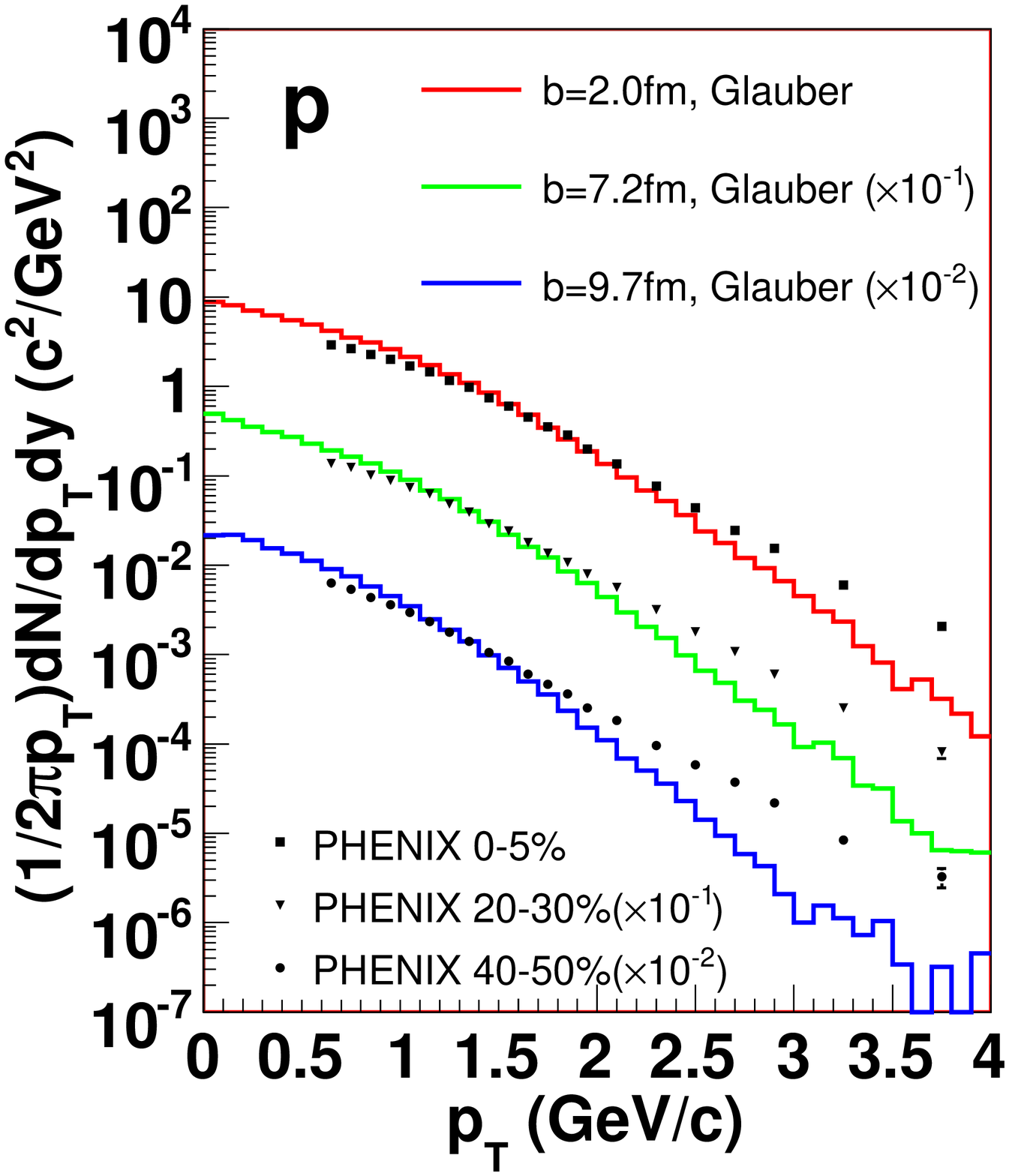}
\caption{(Color online)
  Centrality dependences of the $p_T$ spectra for (a) pions, (b) kaons, 
  and (c) protons obtained from our hydro+cascade hybrid model, compared 
  with data from the PHENIX Collaboration \cite{phenix:pid200} for 
  $200\,A$\,GeV Au+Au collisions. Impact parameters are (from top to 
  bottom) $b\eq2.0$, $7.2$, and $9.7$\,fm, corresponding to the
  0-5\%, 20-30\%, and 30-40\% centrality ranges, respectively.
}
\label{fig:dndpt}
\end{figure*}
%

While these considerations provide a qualitative explanation for the 
harder pion $p_T$-spectrum from the JAM cascade compared to ideal 
hydrodynamics, the same arguments should also hold for protons where
no such effects are seen in Fig.~\ref{fig:dndpt_hydro}(a). As already
stated, this is presumably a consequence of an accidental cancellation 
of delicate thermal and flow effects with viscous corrections for our 
specific choice of $T_\mathrm{th}$ in the hydrodynamic model. Again, a 
full understanding of these results may require comparison with a 
viscous hydrodynamic treatment \cite{Romatschke:2007mq,Song:2007fn}. 

Figure \ref{fig:v2pt_hydro} shows the $p_{T}$ dependence of $v_2$ for 
pions and protons in semi-central Au+Au collisions ($b=7.2$ fm)
at midrapidity ($|\eta|{\,<\,}1.3$), comparing results from the 
hybrid model with ideal hydrodynamics. Whereas, after an initial 
quadratic rise which extends over a larger $p_T$-range for the heavier
protons than the lighter pions \cite{HKHRV01}, the differential elliptic
flow \vtwopt\ from ideal hydrodynamics increases almost linearly with 
$p_{T}$, this increase is tempered in the results from the hadronic
cascade. The differences between the two models is seen to grow with 
increasing $p_T$. Again, this is qualitatively just as expected from 
shear viscous effects \cite{Teaney_visc,Romatschke:2007mq,Song:2007fn}.
Obviously, the different transport properties of the hadronic matter 
in JAM and in hydrodynamics are seen more clearly in the differential 
elliptic flow \vtwopt\ than in the $p_{T}$ spectra.

\subsection{Spectra and elliptic flow for $\pi$, $K$, and $p$}
\label{sec3b}

In this subsection, we compare our results from the hybrid model with
experimental data for identified hadrons. In Fig.~\ref{fig:dndpt}, 
transverse momentum spectra for pions, kaons, and protons from the 
hybrid model are compared with PHENIX data \cite{phenix:pid200}, for
three impact parameters (centrality classes) as shown in the figure.
(The impact parameters are adjusted to give the correct average number 
of participants for each centrality class, as quoted in 
\cite{phenix:pid200}.) In all cases, the experimental data are 
reasonably well reproduced by the hybrid model for low transverse 
momenta to $p_{T} \sim 1.5$-2.0\,GeV/$c$. Additional components (such 
as thermal quark recombination and jet fragmentation, including 
energy loss of fast partons in the fireball medium) would be required
to reproduce the data above $p_{T}\sim 1.5$ GeV/$c$. It should be 
emphasized that, unlike in the purely hydrodynamic approach where
the $p_{T}$ slope is controlled by the choice of kinetic freeze-out 
temperature and the correct hadron yields are ensured by appropriate 
choice of non-equilibrium hadron chemical potentials (see Sec.~\ref{sec2b}),
the hybrid model has no adjustable parameters to reproduce both slope 
and normalization of the transverse momentum spectra. Hadronic cascade 
processes automatically describe both chemical and kinetic freeze-out.

\begin{figure}[ht]
 \includegraphics[bb= 0 0 450 715,width=\linewidth,clip]{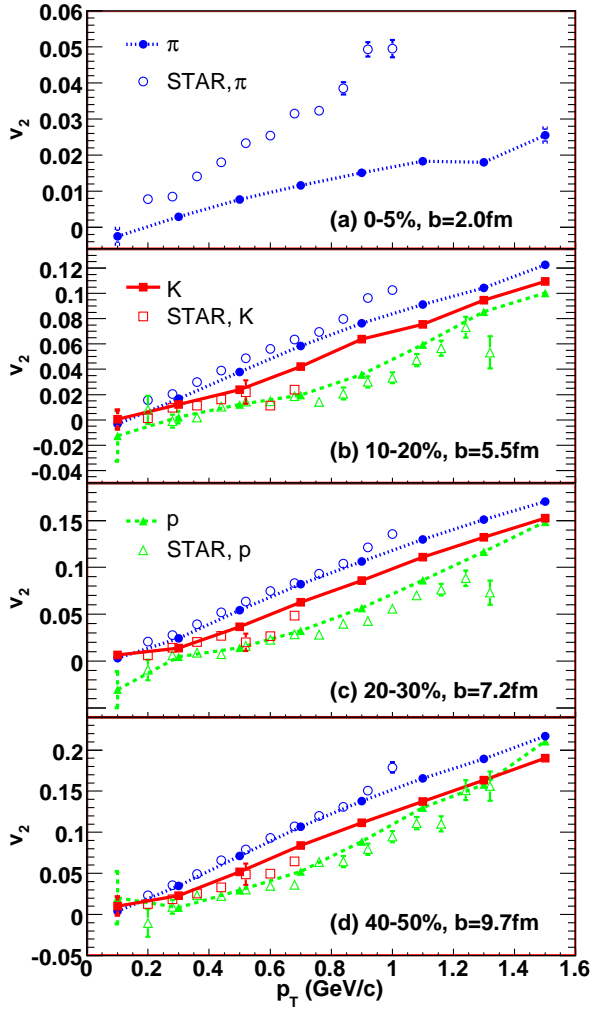}
 \caption{(Color online)
   Transverse momentum dependence of the elliptic flow coefficient $v_2$ 
   for pions (dotted blue), kaons (solid red), and protons (dashed green) 
   from the hybrid model, compared with STAR data for $v_{2}\{2\}$ 
   from 200\,$A$\,GeV Au+Au collisions, in four centrality classes 
   \cite{star:anisotropy}.
 }
 \label{fig:v2pt}
 \end{figure}

In Figure~\ref{fig:v2pt}, we compare the $p_{T}$ dependence of $v_2$
for pions, kaons, and protons with the STAR data for $v_{2}\{2\}$  
\cite{star:anisotropy}, for four centrality classes. For the 
0-5\% centrality class we show only pions since the quality of the
kaon and proton data at this centrality is insufficient for a 
meaningful comparison with theory. The hybrid model correctly describes
the mass ordering of the differential elliptic flow, 
$v_2^\pi(p_T){\,>\,}v_2^K(p_T){\,>\,}v_2^p(p_T)$, as seen in the data
within the low-$p_{T}$ region covered by the figure. Quantitatively, it 
provides a reasonable description up to 50\% centrality, except for 
the most central collisions: Our result for pions at $b\eq2.0$\,fm is
significantly smaller than the data. This can be attributed to the
absence of eccentricity fluctuations in our model calculations 
\cite{Drescher:2006ca,MS03}.

%
%

\begin{figure}[thb]
 \includegraphics[width = \linewidth,clip]{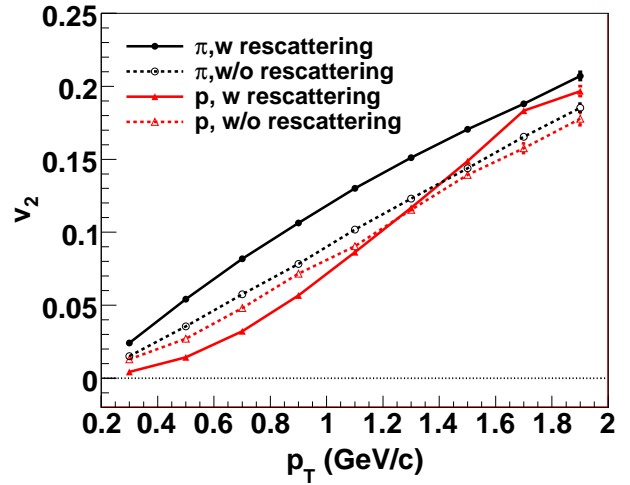}
 \caption{(Color online) 
  Transverse momentum dependence of the elliptic flow parameter
  for pions and protons. Solid (dashed) lines are with (without) 
  hadronic rescattering.
 }
 \label{fig:v2pt_before_after}
 \end{figure}

To better understand the origin of the mass ordering in $v_{2}(p_{T})$,
we compare in Fig.~\ref{fig:v2pt_before_after}, for a selected impact 
parameter of $b\eq7.2$\,fm, the above hybrid model result with a 
calculation where all hadronic rescattering is turned off, allowing
only for decay of the unstable hadron resonances. Whereas just after 
hadronization the differential elliptic flow $v_{2}(p_{T})$ for pions 
and protons looks very similar, the mass splitting gets strongly enhanced
by hadronic rescattering. The smallness of the pion-proton mass splitting 
at $T_\mathrm{sw}$ is partially accidental, because the splitting caused 
by the radial flow already established during the hydrodynamic QGP phase 
\cite{HKHRV01} is significantly decreased by the effect of resonance 
decays which reduces the pion elliptic flow $v_2^\pi(p_T)$ by about 15\%
\cite{Kolb:1999it,Hirano:2000eu}. Hadronic evolution below $T_\mathrm{sw}$ 
steepens the slope of $v_{2}(p_{T})$ for pions \cite{HT02}, due to 
the generation of additional (integrated) $v_{2}$ and the reduction of 
their mean transverse momentum $\langle p_{T} \rangle_\pi$ \cite{HG05}.
(Note that for pions the slope of $v_{2}(p_{T})$ can be simply approximated 
as $dv_{2}(p_{T})/dp_{T}\approx v_2/\langle p_{T} \rangle$ \cite{HG05}.)

%
%

For heavy hadrons, on the other hand, radial flow reduces $v_2$ at low 
$p_{T}$ \cite{HKHRV01}. Assuming positive elliptic flow, 
$v_\perp(\varphi{=}0,\pi) > v_\perp\left(\varphi{=}\frac{\pi}{2}, 
\frac{3\pi}{2}\right)$, the stronger transverse flow $v_\perp$ in the 
reaction plane pushes heavy particles to larger $p_T$ more efficiently 
in the reaction plane than perpendicular to it. In extreme cases 
\cite{HKHRV01} this can, for heavy particles, even lead to a 
{\em depletion} of low-$p_T$ emission into the reaction plane when 
compared with out-of-plane emission, {\it i.e.} to a negative \vtwopt\ 
at low $p_T$ (even though their $p_T$-integrated total elliptic  
flow $v_2$ is positive). But even without going to extremes, this 
mechanism generically reduces \vtwopt\ at low $p_T$ for heavy hadrons.
So it is the generation of additional radial flow \textit{in the 
hadronic stage} which is responsible for (most of) the mass-splitting
of \vtwopt\ observed in the low $p_{T}$ region. 

This mechanism works even if the (extra) radial flow is not perfectly 
hydrodynamic, i.e. if (as is the case in the hadron cascade) the system 
does not remain fully thermalized, with locally isotropic momentum 
distributions. Any type of anisotropic collective transverse motion 
will cause such a mass-splitting of \vtwopt\ at low $p_T$, as long
as the hadron in question participates in the flow. It is worth mentioning 
that in hydrodynamic calculations about half of the final radial flow
in Au+Au collisions at RHIC is generated during the hadronic stage 
(see Fig.~7 in \cite{Kolb:2000sd} and Fig.~5 in \cite{HT02}).
A similar increase in radial flow generated by the JAM cascade is 
documented in Fig.~\ref{fig:dndpt_hydro}(a).

From these observations we conclude that the large magnitude of the 
integrated $v_2$ and the strong mass ordering of the differential 
\vtwopt\ observed at RHIC result from a subtle interplay between 
perfect fluid dynamics of the early QGP stage and dissipative dynamics 
of the late hadronic stage: The large magnitude of $v_2$ is due to 
the large overall momentum anisotropy, generated predominantly in the 
early QGP stage, whereas the strong mass-splitting between the slopes 
of \vtwopt\ at low $p_T$ reflects the redistribution of this momentum 
anisotropy among the different hadron species, driven by the 
continuing radial acceleration and cooling of the matter during the 
hadronic rescattering phase. 

\subsection{Spectra and elliptic flow for $\phi$ mesons}
\label{sec3c}

As noted in Sec.~\ref{sec2d}, $\phi$ mesons (consisting of strange 
quarks) have considerably smaller scattering cross sections in JAM than 
non-strange hadrons \cite{Shor:1984ui}. They are therefore expected to 
show larger dissipative effects in our hybrid model and to not fully 
participate in the additional radial flow generated during the hadronic 
rescattering stage. In kinetic theory language, one expects that the 
$\phi$ mesons decouple from rest of the system earlier than other, 
non-strange hadrons \cite{vanHecke:1998yu}, thereby possibly opening 
a window to extract direct information on collective phenomena in the 
partonic stage from $\phi$-meson spectra \cite{Shor:1984ui}.

To study $\phi$ mesons in our hybrid model we stabilize them by turning 
off their decay channels during the hadronic cascade.
%
 \begin{figure}[th]
 \includegraphics[width = \linewidth,clip]{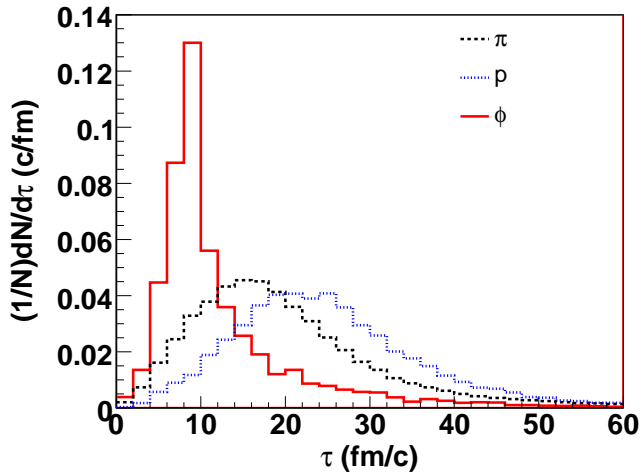}
 \caption{(Color online) Normalized distribution of freezeout times
  for pions (dashed), protons (dotted), and $\phi$ mesons (solid) for 
  $|y|{\,<\,}1$ in Au+Au collisions at $b\eq2.0$\,fm.
 }
 \label{fig:dndt}
 \end{figure}
%

Let us first check how early $\phi$ mesons decouple from the rest of the 
system. Figure~\ref{fig:dndt} shows the normalized distribution of
freeze-out times for pions, protons, and $\phi$ mesons near midrapidity 
$|y|{\,<\,}1$ in central collisions ($\langle b\rangle\eq2.0$\,fm).
Clearly, $\phi$ mesons decouple earlier than pions and protons. The 
freeze-out time distribution for $\phi$ mesons has a prominent peak
at $\tau\eq8$\,fm/$c$, roughly equal to the time of completion of QGP 
hadronization in hydrodynamic simulations. This indicates that only
very few rescatterings happen for $\phi$ mesons during the hadronic 
evolution. Similar results were obtained with the RQMD cascade in 
\cite{vanHecke:1998yu} for $\Omega$ baryons at SPS energies and 
in \cite{CLLSX} for $\phi$ mesons and $\Omega$ baryons at RHIC 
energies. The freeze-out time distributions for pions 
and protons are broadened by both elastic scatterings and resonance 
decays. The long resonance decay tails of the distributions are 
important for interpreting the pion source function that was recently
reconstructed by the PHENIX Collaboration \cite{Adler:2006as} using 
imaging methods.

%
 \begin{figure}[th]
 \includegraphics[bb=10 10 540 620,width=\linewidth,clip]{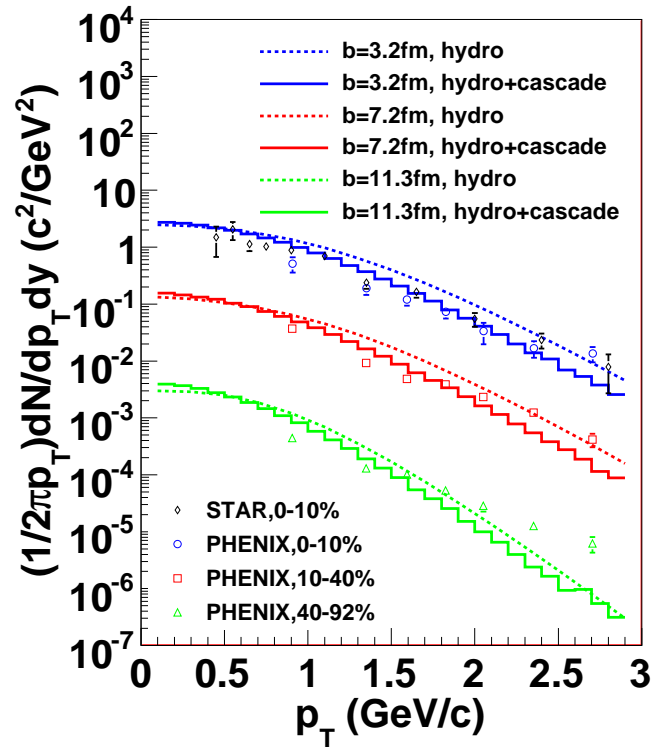}
 \caption{(Color online)  
  Transverse momentum spectra for $\phi$ mesons reconstructed from
  $K^+K^-$ decays in central (blue line), semi-central (red 
  line) and peripheral (green line) Au+Au collisions, compared with 
  PHENIX \cite{phenix:phi} and STAR \cite{star:phi} data. Results from 
  semi-central and peripheral collisions are divided by 10 and 100, 
  respectively. Predictions from ideal hydrodynamics
  with $T_\mathrm{th}\eq100$\,MeV are also shown as dashed lines.
 }
 \label{fig:dndpt_phi}
 \end{figure}
%

In Figure~\ref{fig:dndpt_phi}, $p_{T}$ spectra for $\phi$ mesons from the
hybrid model are compared with PHENIX \cite{phenix:phi} and STAR 
\cite{star:phi} data. Similar to the spectra for pions, kaons, and 
protons in Fig.~\ref{fig:dndpt}, we see good agreement with experiment 
at low $p_T$ ($p_{T}{\,<\,}1.5$\,GeV/$c$). The discrepancy between our 
results and experiment at larger $p_T$ may indicate the appearance of a 
quark-antiquark recombination component in the intermediate $p_{T}$ 
region \cite{Greco:2003xt,Fries:2003vb}. In the presence of such a 
component it is questionable to use the $\phi$-meson spectra over the 
whole available $p_{T}$ region to extract the thermal freeze-out 
temperature and flow for $\phi$ mesons \cite{Schweda:2005sy}; such a 
thermal model fit \cite{Schnedermann:1993ws} should be restricted to 
the region $p_T{\,<\,}1.5$\,GeV/$c$ even if data in that region are 
hard to obtain. 

%
\begin{figure*}[th]
\includegraphics[width=.49\linewidth]{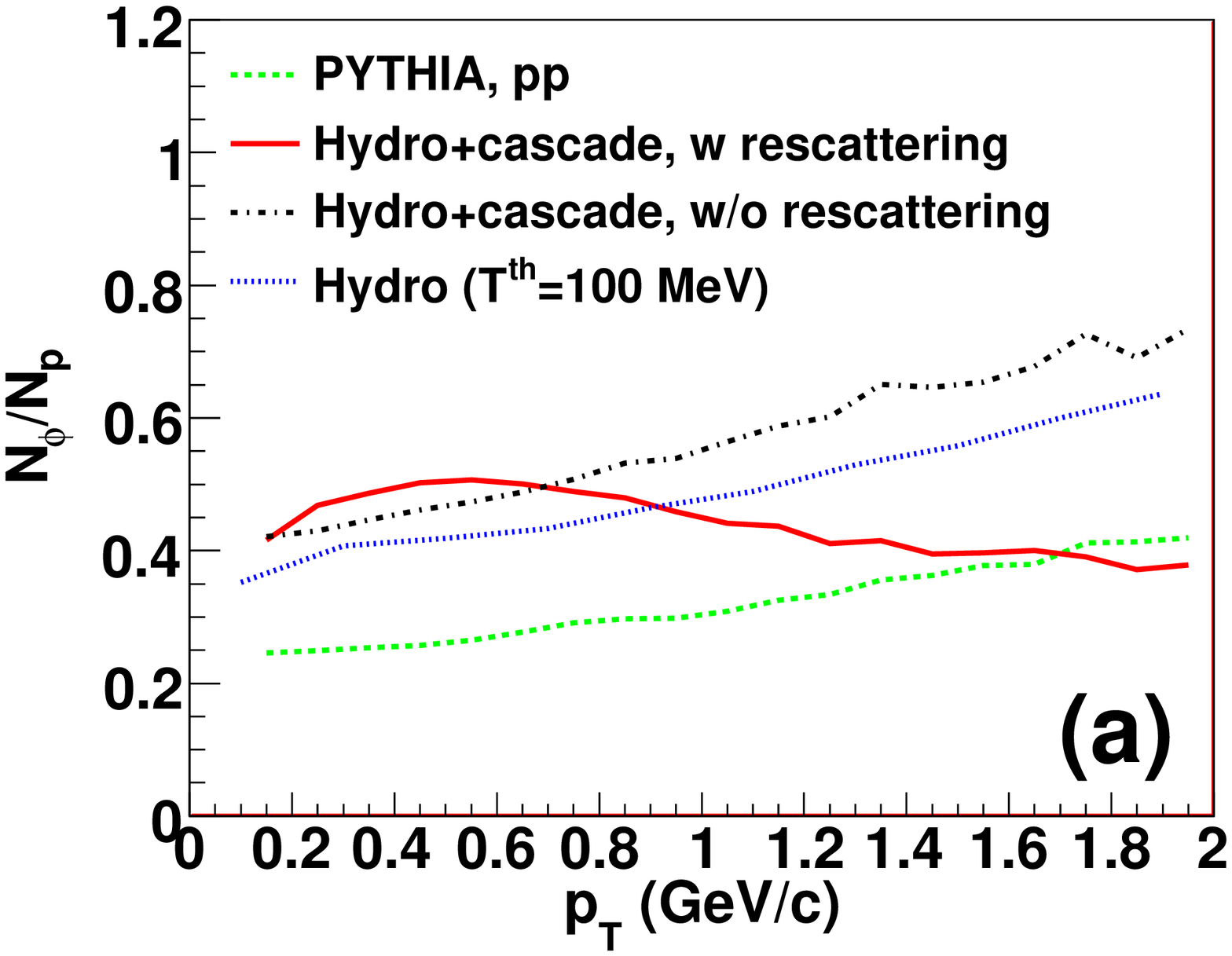}
\includegraphics[width=.49\linewidth]{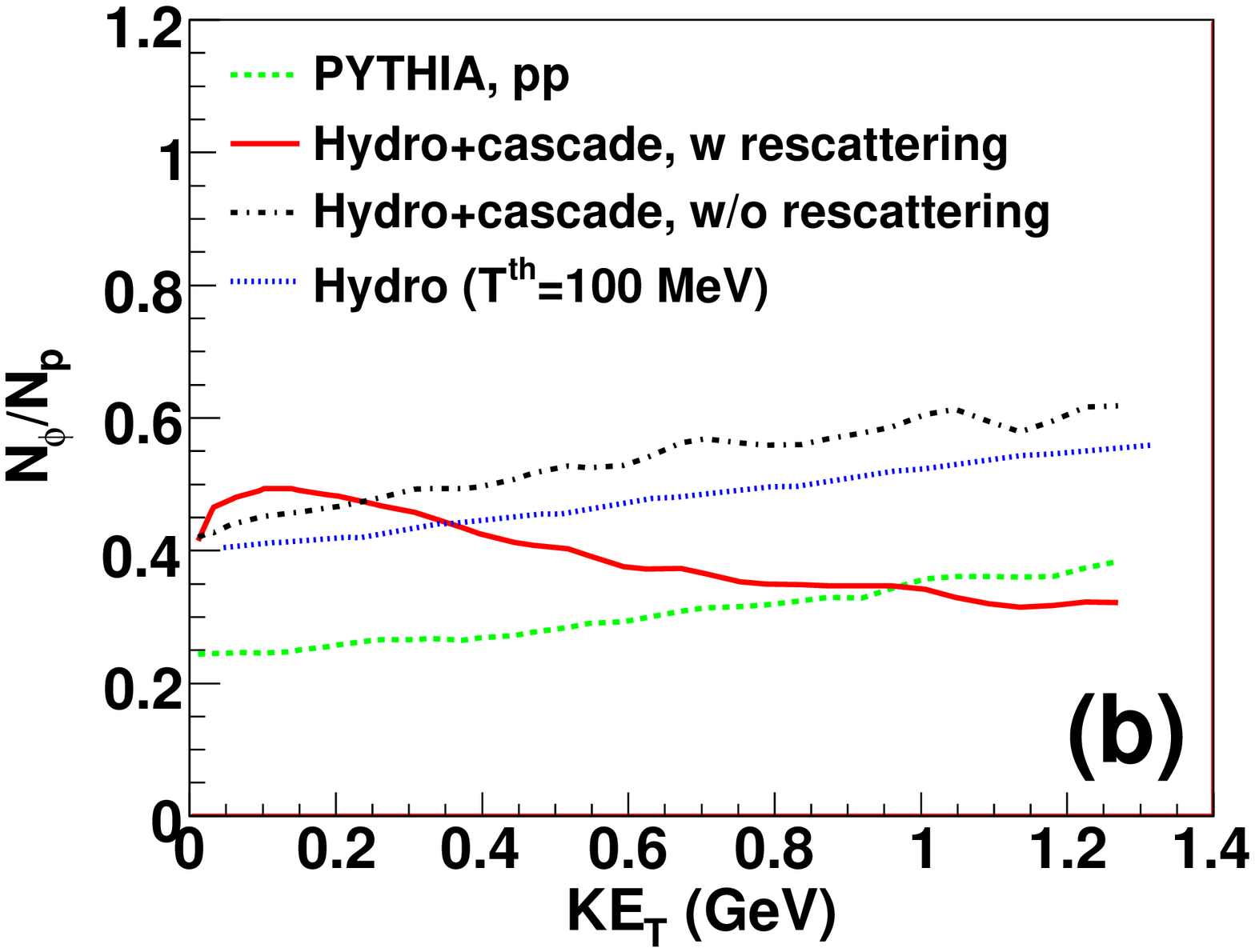}
 \caption{(Color online) 
 The $\phi/p$ ratio as a function of $p_T$
 (left panel) and of transverse kinetic energy 
 $\mathrm{KE}_T{\,\equiv\,}m_T{-}m_0$ (right panel), 
 for different scenarios: central Au+Au collisions in the hybrid model, 
 without hadronic rescattering, and in the
 hydrodynamic model with $T_\mathrm{th}\eq100$\,MeV (dotted). 
 The corresponding ratio for proton-proton collisions (extracted
 from the PYTHIA fit to the experimental data shown in Fig.~\ref{fig:dndpt_pp}
 below) is shown for comparison as the dashed line. See text for
 more discussion.
 }
 \label{fig:ratio}
 \end{figure*}
%

In the hydrodynamic model simulations with $T_{\mathrm{th}} \eq 100$ MeV,
 shown as dashed lines in 
Fig.~\ref{fig:dndpt_phi}, the $\phi$ mesons pick up more additional
radial flow during the hadronic stage,
resulting in flatter $p_T$-spectra
than in the hybrid model and in the data in the low $p_T$ region.
As we 
will show further below, better $\phi$ data at lower $p_T$ and a 
simultaneous analysis of the differential elliptic flow in this region 
should allow to further discriminate between different descriptions 
of the hadronic rescattering stage.   

The effects of radial flow, and the difference in how additional radial
flow generated during the hadronic rescattering stage is picked up by
protons and $\phi$ mesons (which have rather similar masses), can be
enhanced by studying the $p_T$ or transverse kinetic energy dependence 
of the $\phi/p$ ratio. A thermalized medium {\em without} radial flow
features $m_T$-scaling, {\it i.e.} all $m_T$-spectra have identical
slopes, and for such a static fireball the $\phi/p$ ratio, when plotted 
as a function of transverse kinetic energy 
$\mathrm{KE}_T{\,\equiv\,}m_T{-}m_0$, would be a constant horizontal line.
For a thermalized {\em expanding} medium, $m_T$-scaling is broken by
radial flow (which couples differently to particles with different
masses), resulting in a non-zero slope of the ratio $\phi/p(\mathrm{KE}_T)$. 
Perhaps somewhat counterintuitively, this slope of the $\phi/p$ {\em ratio} 
does not grow monotonically with the radial flow $v_\perp$ but, after an 
initial rise, decreases again when the flow becomes so large that the 
hadron $m_T$-spectra become very flat; in the limit of ``infinite flow'' 
({\it i.e.} $\gamma_\perp\eq1/\sqrt{1{-}v_\perp^2}{\,\to\,}\infty$) the 
hadron $m_T$-spectra, and thus their ratios, become again perfectly flat. 
   
In Figure~\ref{fig:ratio} we show the $\phi/p$ ratio, both as a function of 
transverse kinetic energy (right panel) and of $p_T$ (left panel). 
It should be noted here that weak decay contribution is not included
in proton yields.
In the 
latter case the connection to radial flow is less straightforward, since
the kinematics of the transformation from $m_T$ to $p_T$ depends on
mass and introduces additional growth with $p_T$ for the ratio. In
both representations one sees, however, by comparing the curves for the
hydro+cascade model without rescattering (corresponding to ideal hydrodynamics
with $T_{\mathrm{th}}\eq169$\,MeV) and for the ideal hydrodynamic model with 
$T_\mathrm{th}\eq100$\,MeV, that (i) the ratio increases with $p_T$ or
$\mathrm{KE}_T$ due to radial flow effects, and that (ii) the rate of
increase drops when the freeze-out temperature $T_\mathrm{th}$ is decreased,
due to build-up of additional radial flow. 
Surprisingly, the ratio increases
even 
 in $pp$ collisions, but for entirely different reasons,
unrelated to collective flow: The $\phi$ spectrum from $pp$ collisions 
shown in Fig.~\ref{fig:dndpt_pp} below is considerably flatter than the 
proton spectrum, leading to the prominent rise of the $\phi/p$ ratio
with $p_T$. The most interesting feature of Fig.~\ref{fig:ratio} is 
 that the $\phi/p$ ratio from the hybrid model does not at all
increase with $p_T$ or $\mathrm{KE}_T$ (except at very low 
$p_T{\,<\,}500$\,MeV/$c$). Instead, it {\em decreases} over almost
the entire range of transverse kinetic energy shown in the figure. This 
decrease is due to the flattening of the proton spectrum by
hadronically generated radial flow in which the weakly coupled $\phi$ 
mesons do not participate. The comparison with $pp$ collisions and 
hydrodynamic model simulations in Fig.~\ref{fig:ratio} shows that the 
observation of such a decreasing $\phi/p$ ratio would be an unambiguous
signature for early decoupling of $\phi$ mesons from the hadronic rescattering
dynamics.

%
\begin{figure*}[th]
\includegraphics[width=.32\linewidth]{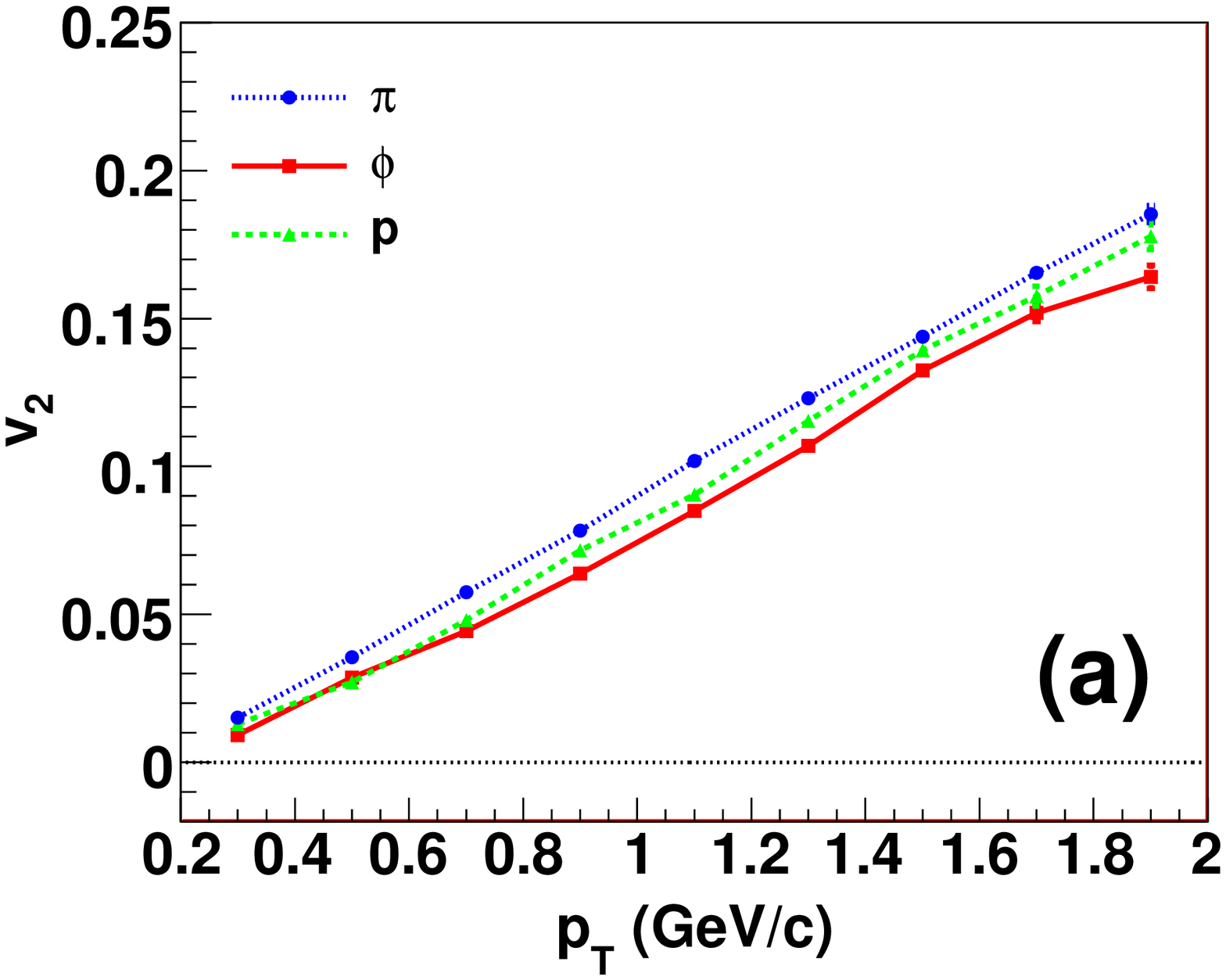}
\includegraphics[width=.32\linewidth]{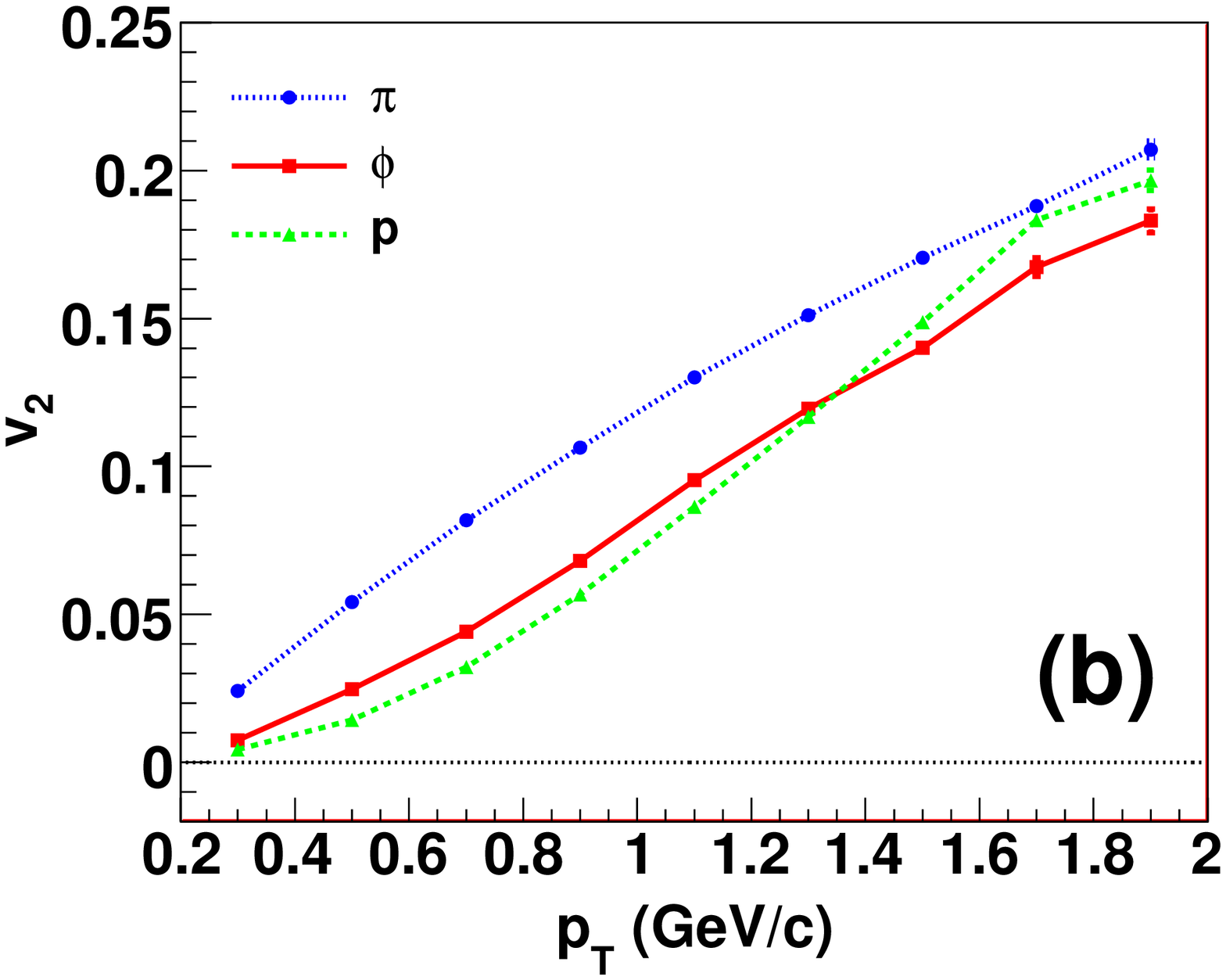}
\includegraphics[width=.32\linewidth]{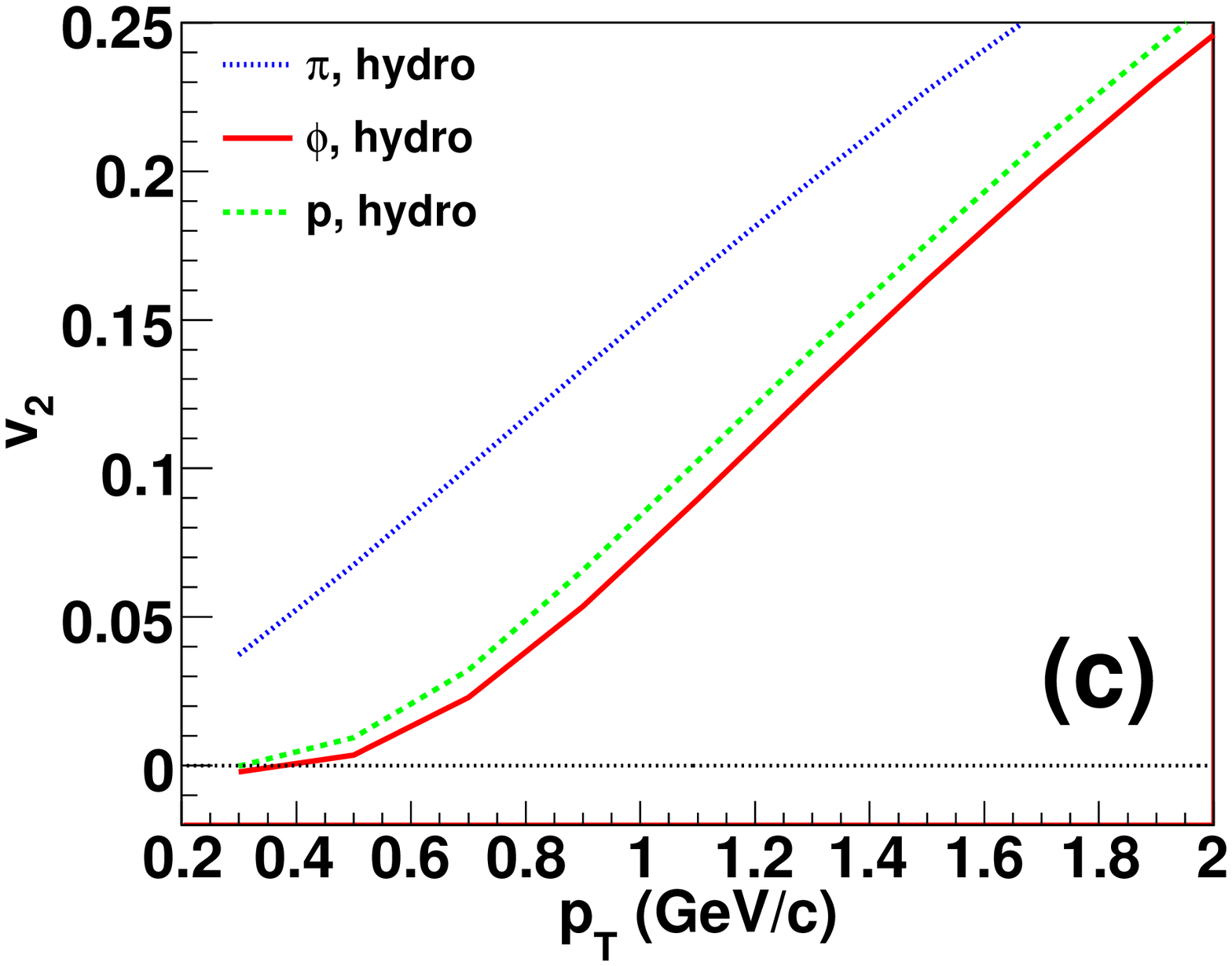}
 \caption{(Color online) Transverse momentum dependence of the elliptic 
  flow parameters for pions (dotted blue), protons (dashed green),
  and $\phi$ mesons (solid red), for Au+Au collisions at $b\eq7.2$\,fm.
  (a) Before hadronic rescattering. (b) After hadronic rescattering.
  (c) Ideal hydrodynamics with $T_\mathrm{th}\eq100$\,MeV.
  The results for pions and protons are the same as shown in 
  Fig.~\ref{fig:v2pt_before_after}.
 }
 \label{fig:v2pt_before_phi}
 \end{figure*}
%

We now proceed to the discussion of dissipative effects during the hadronic 
rescattering stage on the differential elliptic flow \vtwopt. Figure 
\ref{fig:v2pt_before_phi} shows \vtwopt\ from the hybrid model for 
$\pi$, $p$, and $\phi$. We consider semi-central collisions (20-30\% 
centrality), choosing impact parameter $b\eq7.2$\,fm. In the absence of 
hadronic rescattering we observe the hydrodynamically expected mass ordering 
$v_2^\pi(p_T){\,>\,}v_2^p(p_T){\,>\,}v_2^\phi(p_T)$ 
(Fig.~\ref{fig:v2pt_before_phi}(a)), but just as in 
Fig.~\ref{fig:v2pt_before_after} (dashed lines) the mass splitting is 
small. Figure~\ref{fig:v2pt_before_phi}(b) shows the effects of hadronic 
rescattering: while the \vtwopt\ curves for pions and protons separate 
as discussed before (at low $p_T$ the pion curve moves up while the proton 
curve moves down), \vtwopt\ for the $\phi$ meson remains almost unchanged 
\cite{Chen:2006ub}. As a result of rescattering the proton elliptic 
flow ends up being smaller than that of the $\phi$ meson, 
$v_2^p(p_T){\,<\,}v_2^\phi(p_T)$ for $0{\,<\,}p_T{\,<\,}1.2$\,GeV/$c$, 
even though $m_\phi{\,>\,}m_p$. Hadronic dissipative effects are seen 
to be particle specific, depending on their scattering cross sections 
which couple them to the medium. The large cross section difference 
between the protons and $\phi$ mesons in the hadronic rescattering 
phase leads to a violation of the hydrodynamic mass ordering at low 
$p_T$ in the final state.

This is the most important new result of our work. Current experimental
data \cite{Afanasiev:2007tv,Abelev:2007rw} neither confirm nor contradict 
this predicted behavior, due to the difficulty of reconstructing low-$p_T$ 
$\phi$ mesons from their decay products. 
If it turns out that
high precision $\phi$-meson $v_2$ data at low $p_T$
 show violation of mass ordering,
it will be evidence
for strong momentum anisotropy having developed already during the QGP
stage, with the contribution carried by $\phi$ mesons not being
redistributed in $p_T$ by late hadronic rescattering.
At intermediate $p_T$, recent 
data \cite{Afanasiev:2007tv,Abelev:2007rw} confirm the prediction from 
the quark coalescence model \cite{Molnar:2003ff,Nonaka:2003hx} that there 
the elliptic flow should scale with the number of constituent quarks:
$v_2^\phi(p_T)\approx v_2^{\pi,K}(p_T)\approx\frac{2}{3}v_2^p(p_T)$,
in spite of the similar $\phi$ and $p$ masses which are much larger than
those of the pions and $K$ mesons. We hope that the present paper motivates 
an effort to extend these data to lower $p_T$ in order to test our prediction
here that, at low $p_T$, $v_2^p(p_T){\,<\,}v_2^\phi(p_T)$ in spite of 
$m_\phi{\,>\,}m_p$. While the former observation suggests that at 
intermediate $p_T$ ($2\,\mathrm{GeV}/c{\,<\,}p_T{\,<\,}6$\,GeV/$c$) quark
coalescence during the quark-hadron phase transition controls the finally
observed elliptic flow of all hadrons, without measurable distortion by 
subsequent hadronic reinteractions, confirmation of our prediction would
confirm the importance of hadronic rescattering on low-$p_T$ hadrons,
with results that depend on the magnitude of the scattering cross sections 
of the various hadron species.

We close this section with a discussion of the implications of our 
hybrid model results for the {\em nuclear modification factor} 
\begin{eqnarray}
\label{eq:raa}
  R_{AA}(p_{T}) = 
  \frac{\frac{dN_{AA}}{p_{T}dp_{T}dy}}
       {N_{\mathrm{coll}}\,\frac{dN_{pp}}{p_{T}dp_{T}dy}}
  = \frac{\frac{dN_{AA}}{p_{T}dp_{T}dy}}
         {T_{AA}\,\frac{d\sigma_{pp}}{p_{T}dp_{T}dy}}.
\end{eqnarray}
The observed suppression of pion yields at intermediate to high $p_{T}$
\cite{experiments} provides evidence of jet quenching in relativistic 
heavy-ion collisions. For baryons, this suppression effect is counteracted
in the intermediate $p_T$ region by collective flow effects which, at low 
$p_T$, lead to a rise of the $p/\pi$ (or, more generally, heavy/light) 
ratio as a function of $p_T$. Collective flow effects extend into the 
intermediate $p_T$ region $2\,\mathrm{GeV}/c{\,<\,}p_T{\,<\,}6$\,GeV/$c$ 
even though the hydrodynamic picture 
%
 \begin{figure}[h]
 \includegraphics[width = \linewidth,clip]{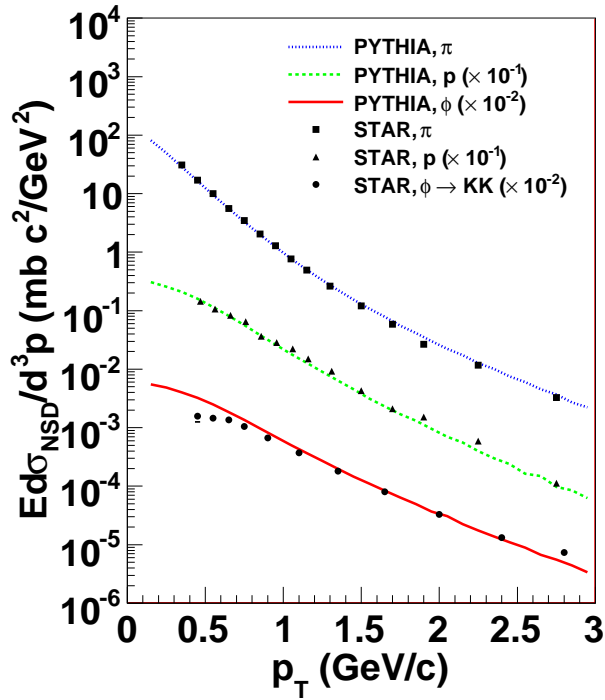}
 \caption{(Color online) 
 Invariant cross sections as a function of $p_{T}$ in non-singly 
 diffractive $pp$ collisions for pions, protons, \cite{star:pp_pikp} 
 and $\phi$ mesons. Dotted, dashed, and solid lines are results from PYTHIA
 for pions , protons,
  and $\phi$ mesons, 
 respectively. 
 }
 \label{fig:dndpt_pp}
 \end{figure}
%
%
 \begin{figure*}[t]
 \includegraphics[width = 0.49\linewidth,clip]{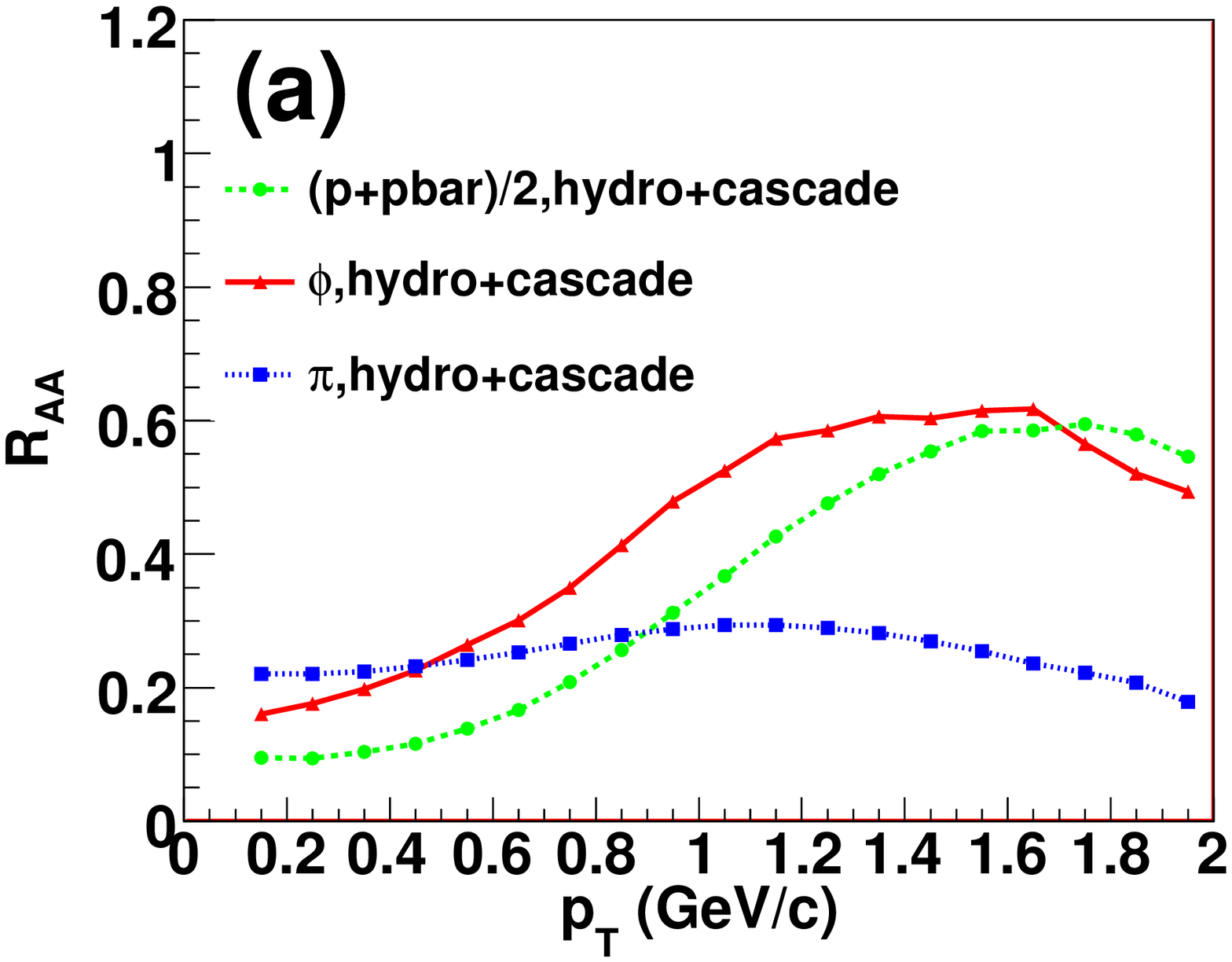}
 \includegraphics[width = 0.49\linewidth,clip]{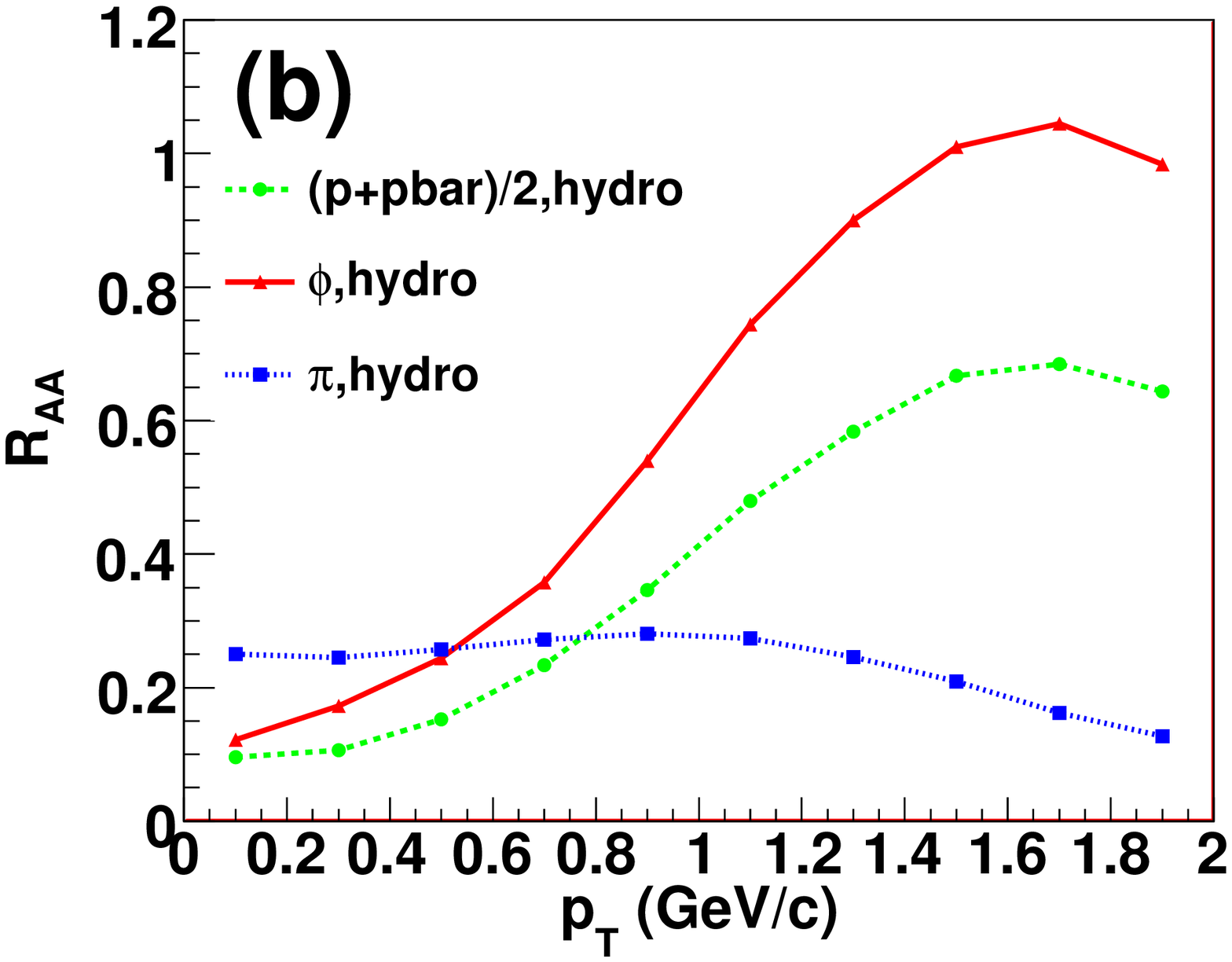}
 \caption{(Color online)
  Nuclear modification factors $R_{AA}$ for pions (blue), $\phi$ mesons 
  (red), and protons/antiprotons (green), for Au+Au collisions at 
  $b=3.2$ fm (corresponding to 0-10\% centrality). Shown are
  predictions from (a) the hybrid model and (b) from ideal hydrodynamics
  with $T_\mathrm{th}\eq100$\,MeV.
 }
 \label{fig:raa}
 \end{figure*}
%
is known to gradually break down
above $p_T{\,>\,}1.5-2.5$\,GeV/$c$ \cite{Heinz:2004ar}. Quark coalescence 
is one of the key mechanisms by which low-$p_T$ collectivity on the
quark-gluon level is transferred to the hadron spectra at intermediate
$p_T$ during the hadronization process \cite{Greco:2003xt,Fries:2003vb,%
Molnar:2003ff}, leading to (unsuppressed) values of $R_{AA}$ (or of
$R_{\mathrm{CP}}$, the ratio of yields per number of binary collisions 
in central and peripheral collisions) of order unity for baryons at 
$p_{T}{\,\sim\,}2$-3 GeV/$c$ \cite{experiments,Abelev:2007rw}. We will
show that hadronic rescattering following QGP hadronization affects 
$R_{AA}$ at low $p_T$ instead.

The PHENIX \cite{phenix:phi} and STAR \cite{Abelev:2007rw} Collaborations
have recently measured $R_{\mathrm{CP}}$ for $\phi$ mesons. The PHENIX 
data show a suppression of $\phi$ mesons by about a factor 2 (with 
relatively large error bars) in the region 
$1\,\mathrm{GeV}/c{\,<\,}p_{T}{\,<\,}3$\,GeV/$c$, consistent with that of 
pions, while protons and antiprotons are unsuppressed \cite{phenix:phi}.
This seems to be in contradiction with collective flow arguments which 
predict $R_\mathrm{CP}^\phi{\,>\,}R_\mathrm{CP}^p$ since 
$m_{\phi}{\,>\,}m_{p}$, but consistent with the valence quark scaling 
predicted by the quark coalescence model {\cite{Greco:2003xt,Fries:2003vb,%
Molnar:2003ff}. The more recent and precise STAR data \cite{Abelev:2007rw}, 
on the other hand, show an $R_\mathrm{CP}$ 
for $\phi$ mesons that follows the one for pions and exceeds the one
for protons for $p_T{\,<\,}1$\,GeV/$c$, but then follows the rise of 
the proton $R_\mathrm{CP}$ above the pion one for $p_T{\,>\,}1$\,GeV/$c$, 
lagging only slightly behind the protons and reaching a value halfway 
between pions and protons in the region $p_T{\,\sim\,}2{-}3$\,GeV/$c$ 
where $R_\mathrm{CP}^p$ peaks at a value of $\sim 1$.

Given this somewhat contradictory experimental situation, we offer
a prediction from our hybrid model (cautioning beforehand that this 
model does not include any quark-recombination contributions which 
are expected to become important above $p_T{\,\gtrsim\,}1.5{-}2$\,GeV/$c$)
in Figure~\ref{fig:raa}. To construct this Figure, we first fitted
the experimentally measured $p_{T}$-spectra for pions and protons 
\cite{star:pp_pikp} as well as for $\phi$ mesons \cite{star:pp_phi}
in non-singly diffractive (NSD) $pp$ collisions ({\it i.e.} inelastic 
collisions excluding single diffractive events). The fit, shown in
Fig.~\ref{fig:dndpt_pp}, is performed with the help of the event 
generator PYTHIA 6.403 \cite{Sjostrand} which, once properly tuned, 
yields smooth reference $p_{T}$-spectra for $pp$ collisions. PYTHIA 
is based on leading order perturbative QCD for semi-hard processes 
combined with a Lund string fragmentation scheme for soft particles. 
It works quite well for pions, protons and $\phi$ mesons
 with default parameters 
\cite{Sjostrand}, except for a necessary readjustment of the $K$ factor 
to $K\eq1.8$. 
We note that here exceptionally
this comparison includes all resonance decays including weak ones since
the STAR data show the inclusive spectra.
We take the resulting spectra as our
$pp$ reference, after removing
weak decay contributions and multiplying them with the ratio 
$\sigma_{\mathrm{in}}/\sigma_{\mathrm{NSD}}$ to correct for the NSD 
trigger. For the required cross sections PYTHIA provides the estimates 
$\sigma_{\mathrm{NSD}}\eq32$ mb and $\sigma_{\mathrm{in}}\eq42$ mb.

With these reference spectra the nuclear modification factors $R_{AA}$ 
can now be calculated from the results shown in Figs.~\ref{fig:dndpt_hydro},
\ref{fig:dndpt} and \ref{fig:dndpt_phi}. For pions, protons, and $\phi$ 
mesons they are shown as functions of $p_T$ in Fig.~\ref{fig:raa}, for 
Au+Au collisions at impact parameter $b\eq3.2$\,fm ({\it i.e.} 0-10\% 
centrality). Figure~\ref{fig:raa}(a) shows the predictions for the 
hybrid model. While for pions $R_{AA}(p_T)$ is almost flat, 
$R_{AA}^\pi{\,\sim\,}0.15{-}0.25$, the $R_{AA}(p_T)$ curves for protons and 
$\phi$ mesons increase with $p_{T}$ as expected from radial flow arguments 
(radial flow hardens the $p_{T}$ spectra for heavy particles). 
The rate of increase for the $\phi$ mesons
is very similar to that for protons, 
culminating in a peak value of ${\sim\,}60\%$ at 
$p_T{\,\sim\,}1.2-1.4$\,GeV/$c$ for $\phi$'s whereas the $R_{AA}$ for
protons peaks at a value of ${\sim\,}60\%$ near 
$p_T{\,\sim\,}1.8$\,GeV/$c$. Figure~\ref{fig:raa}(b) shows the 
corresponding curves for the ideal fluid dynamical simulation with
$T_\mathrm{th}\eq100$\,MeV. For pions and protons, the differences to
the hybrid model are minor (at least in the $p_T$ range covered in
the Figure), reiterating the observation made in connection with 
Fig.~\ref{fig:dndpt_hydro} that the buildup of additional radial
flow during the hadronic stage is similar in both models and viscous
effects become clearly recognizable only at larger $p_T$. For $\phi$
mesons one observes a much faster rise of $R_{AA}(p_T)$ in the 
hydrodynamic approach, resulting in a larger peak value of 
${\sim\,}105\%$ at a larger $p_T$ value (${\sim\,}1.7$\,GeV/$c$)
than for the hybrid model. The reason for these effects is obviously
the larger amount of radial flow picked up during the hadronic stage
in the hydrodynamic model and the resulting hardening of the $\phi$
spectrum. The much weaker rise of $R_{AA}^\phi(p_T)$ in the hybrid model 
can thus be traced directly to the lack of $\phi$ meson rescattering 
during the hadronic stage. 

We note that, even in the hydrodynamic model, the nuclear modification 
factor $R_{AA}(p_T)$ doesn't show a monotonic mass-ordering at low 
$p_T$. Naive expectations based on the mass-ordering of the spectral
slopes (which reflect radial flow effects) are invalidated by the 
fact that the $\phi$ $p_T$-spectra from $pp$ collisions are flatter 
than the corresponding proton spectra. Since these spectra enter the
denominator of $R_{AA}$, they distort its $p_T$ dependence differently
for protons and $\phi$ mesons.

We also comment that at $p_T \sim 2$\,GeV/$c$,
 the characteristics of the observed mass-scaling 
violation in Fig.~\ref{fig:raa}(a)
are qualitatively similar to those expected (and observed) 
in the quark coalescence picture at intermediate $p_T$ 
($2\,\mathrm{GeV}/c{\,<\,}p_T{\,<\,}6$\,GeV/$c$) \cite{Nonaka:2003hx}. 
The differences are quantitative: our prediction for $R_{AA}$ features 
neither a monotonic mass-ordering at low $p_T$ nor the strict valence 
quark scaling predicted by the quark-coalescence picture at intermediate 
$p_T$.

\section{Conclusions}
\label{sec4}

We have studied effects of hadronic dissipation on the spectra, 
differential elliptic flow, and nuclear modification factor of pions,
kaons, protons, and $\phi$ mesons from Au+Au collisions at RHIC, using 
a hybrid model which treats the early QGP phase macroscopically as a 
perfect fluid and the late hadronic phase microscopically with a hadronic 
cascade. For transverse momenta below $1.5$\,GeV/$c$ and not too peripheral
collisions, the hybrid model gives a reasonable description of the 
measured pion, kaon, proton and $\phi$ meson $p_T$-spectra. In peripheral 
collisions ($b\eq9$\,fm and larger) the model spectra tend to be somewhat
steeper than measured. The centrality dependence of the differential 
elliptic flow \vtwopt\ of pions, kaons and protons is better described 
by the hybrid model than in a purely hydrodynamic approach. 

For pions, kaons, and protons, which have relatively large scattering 
cross sections, hadronic rescattering is seen to generate additional 
collective transverse flow, but not so for the much more weakly interacting 
$\phi$ mesons. However, even for pions and protons the extra hadronic 
transverse flow effects are not ``ideal'' but exhibit obvious viscous 
features: Their $p_T$-spectra are hardened while the growth of their 
elliptic flow \vtwopt\ with increasing $p_T$ is tempered by viscous 
corrections whose importance is in both cases observed to increase with 
transverse momentum. The well-known mass-splitting of the differential 
elliptic flow \vtwopt\ observed in hydrodynamic models is seen to be 
mostly generated during the hadronic rescattering phase and to be largely 
due to a {\em redistribution} of the momentum anisotropy built up during 
the QGP stage. This redistribution is caused by the mass-dependent 
flattening of the transverse momentum spectra by additional radial flow 
generated during the hadronic stage. The much more weakly 
interacting $\phi$ mesons do not participate in this additional radial 
flow and thus are not affected by this redistribution of momentum 
anisotropies: their differential elliptic flow remains almost unaffected 
by hadronic rescattering. The net result of dissipative hadronic 
rescattering is therefore that the differential elliptic flow \vtwopt\ 
of protons {\em drops below} that of the $\phi$ mesons, in violation of 
the hydrodynamic mass-ordering. A similar violation of the mass-ordering
is seen in the nuclear modification factor $R_{AA}(p_T)$ at 
$p_T \sim 2$ GeV/$c$
where, after hadronic rescattering, the curve for $\phi$ mesons ends up 
{\em between} those for pions and protons even though the $\phi$ is 
heavier than both of them. For the $\phi/p$ ratio, the lack of interaction
between the $\phi$ mesons and its accelerating hadronic environment should 
manifest itself in an unexpected but unambiguous {\em decrease} with 
increasing transverse kinetic energy.

The results presented here underscore the conclusion of Ref.~\cite{HHKLN}
that hadronic dissipation may be very important at RHIC and at lower
beam energies and should be properly accounted for in attempts to 
quantitatively account for the experimental data collected from heavy-ion 
collisions. With \vtwopt\ and $R_{AA}(p_T)$ for low-$p_T$ $\phi$ mesons 
and the dependence of the $\phi/p$ ratio on $p_T$ or transverse kinetic
energy $\mathrm{KE}_T$, we have identified three additional critical 
observables which should be helpful in sorting out the interplay between 
hydrodynamic evolution during the early QGP stage and dissipative hadronic 
expansion during the late stage of the hot and dense fireballs created in 
these collisions. An accurate extraction of the value for the specific 
shear viscosity $\eta/s$ of the QGP created at RHIC requires a proper 
accounting for effects from late hadronic viscosity. Here, an attempt 
has been made to do this, by coupling the hydrodynamic model to a 
hadronic cascade.

\acknowledgments
\vspace*{-2mm}
 This work was supported by the U.S. DOE under contracts 
 DE-FG02-01ER41190 (U.H.),
 DE-AC02-98CH10886 (D.K.)
 and DE-FG02-87ER40331.A008 (R.L.).
 The work of T.H. was partly supported by
 Grant-in-Aid for Scientific Research
 No.~19740130.  



\begin{thebibliography}{99}
\bibitem{fn1}
  Note that the authors of \cite{NonakaBass} use different algorithms
  for both the hydrodynamic and hadronic cacade codes than employed 
  here and in \cite{HHKLN}.
\bibitem{fn2}
  The main difference is in baryon-antibaryon annihilation channels 
  where the hadronic cascade models lack a consistent description of  
  inverse processes involving more than two colliding particles. This
  lack of detailed balance leads to an excess loss of baryons and 
  antibaryons during the hadronic rescattering phase \cite{BassDumitru,RS01}.
\end{thebibliography}

\end{document}